\documentclass[aps,prb,superscriptaddress,twocolumn,longbibliography]{revtex4-2}
\usepackage{bbm}
\usepackage{graphicx}
\usepackage{dcolumn}
\usepackage{bm}
\usepackage{subfigure}
\usepackage{amsmath}
\usepackage{amssymb}
\usepackage{feynmf}
\usepackage{hyperref}
\usepackage{color}
\usepackage{braket}
\usepackage{ulem}
\usepackage{xcolor}

\usepackage{attachfile}

\newcommand{\bk}{\boldsymbol k}

\newcommand{\zb}{\color {black}}
\usepackage{times}

\begin{document}

\title{{\zb Large-Chern-number flat bands, anomalous Dirac cones, and unconventional superfluidity \\in square-lattice systems 
with SU($N$) non-Abelian  gauge fields}}
	
\author{Wentao Zhong}
\affiliation{Guangdong Provincial Key Laboratory of Magnetoelectric Physics and Devices,
State Key Laboratory of Optoelectronic Materials and Technologies, School of Physics, Sun Yat-Sen University, Guangzhou 510275, China}
	
\author{Zhongbo Yan}
\email{yanzhb5@mail.sysu.edu.cn}
\affiliation{Guangdong Provincial Key Laboratory of Magnetoelectric Physics and Devices, State Key Laboratory of Optoelectronic Materials and Technologies, School of Physics, Sun Yat-Sen University, Guangzhou 510275, China}

\date{\today}
	
\begin{abstract}
We study topological band structures and superfluid phases in two-dimensional square-lattice systems with homogeneous SU($N$) non-Abelian gauge fields. Starting from an SU(4) gauge-field model related to the Hofstadter model with flux $\alpha=1/4$, we show that the lowest and highest bands are isolated Chern bands that carry Chern numbers {\zb $C=-4$ and give rise to four} chiral edge modes in a strip geometry. {\zb Remarkably,
although the two middle bands touch and form $16$ gapless Dirac cones, their combined Chern number is $C=8$}. We {\zb then} generalize the construction to SU($N$) systems and reveal an even--odd structure of the band topology: {\zb when $N$ is odd, all bands are isolated and carry nonzero Chern numbers; 
 when $N$ is even, the two middle bands touch at $N^{2}$ Dirac points, while all other bands remain isolated and topologically nontrivial}.  
 We find that the uppermost and lowermost bands become increasingly flat and their Berry curvature becomes more uniform as $N$ increases, providing a promising {\zb platform for realizing} fractional Chern insulating phases. {\zb We further examine the spin-$3/2$ SU(4) model with on-site attractive Hubbard interactions, exploring its superfluid phases at partial filling}.
We find that the non-Abelian gauge field breaks the hidden SO(5) degeneracy of {\zb the} quintet pairing and selects distinct nematic superfluid states. {\zb For fillings} in the middle-band regime, the resulting spectrum can host topologically protected Bogoliubov Fermi surfaces. Our results provide a starting point for exploring both topological band physics and unconventional superfluidity in synthetic SU($N$) cold-atom systems.
\end{abstract}
	
\maketitle
	
\section{Introduction}

Topological phases of matter have profoundly changed our understanding of quantum states. Unlike conventional phases characterized by local order parameters and symmetry breaking, topological phases are distinguished by global invariants {\zb encoded in the wave functions of the bulk Hamiltonian}~\cite{Hasan2010,Qi2011,Chiu2016,Xiao2010,kane2005,bernevig2006}. In two-dimensional systems, the Chern number provides a paradigmatic example: it determines quantized Hall responses and guarantees robust chiral boundary modes through the bulk-boundary correspondence~\cite{thouless1982,halperin1982,haldane1988,Hatsugai1993}. These properties make topological states not only conceptually important, but also potentially useful for realizing dissipationless transport and fault-tolerant quantum computation~\cite{Nayak2008}. A central goal in this field is therefore to find highly controllable platforms in which topological band structures and their interaction-driven descendants can be engineered, manipulated, and probed in a clean and tunable way~\cite{bloch2008,lewenstein2007,bloch2012,gross2017}. 

Among the available quantum-simulation platforms, ultracold atomic gases are particularly attractive because of their high controllability and cleanliness. Optical lattices, Raman couplings, laser-assisted tunneling, and lattice modulation techniques allow one to engineer synthetic gauge fields and spin-orbit couplings {\zb with tunable forms that are difficult to achieve} in solid-state materials~\cite{lin2009,lin2009Nature,lin2011,Dalibard2011,goldman2014,cooper2019,galitski2013,wang2012,zhai2015,cheuk2012,aidelsburger2011,aidelsburger2013,Goldman2016}. This is especially relevant for topological band physics, since spin-orbit coupling provides {\zb a natural} mechanism for {\zb locking} internal degrees of freedom to orbital motion and generating nontrivial band geometry. In solid-state systems, however, the form of spin-orbit coupling is {\zb stringently constrained} by microscopic crystal symmetries and by the properties of real electron spin. By contrast, synthetic spin-orbit coupling in cold atoms can be designed more flexibly. Moreover, cold atoms naturally possess multiple internal hyperfine states originating from the coupling between electronic and nuclear angular momenta, making them well suited for studying high-spin and multicomponent quantum systems~\cite{Wu2003,Wu2006,Ho1999,Yip1999,gorshkov2010,taie2012,cazalilla2014,zhang2014,pagano2014,scazza2014,Taie2022,Pasqualetti2024}. These internal states can be viewed as a synthetic spin space, on which non-Abelian gauge potentials act as matrix-valued hopping phases. As a result, synthetic spin-orbit coupling can be generalized from the conventional SU(2) form to SU($N$) gauge fields for multicomponent atoms. Previous work has introduced SU($N$) spin-orbit coupling in ultracold atoms and demonstrated that a homogeneous SU(3) non-Abelian gauge field on a square lattice can realize topologically nontrivial Chern bands, whereas the corresponding SU(2) {\zb gauge field} cannot~\cite{Barnett2012}. 
{\zb This distinction arises because the uniform SU(2) gauge field, which reduces to conventional spin-orbit coupling, preserves time-reversal symmetry (TRS), while its SU(3) counterpart breaks it, thereby giving rise to nonzero Chern numbers~\cite{Bornheimer2018}. It is important to note that,} in synthetic spin systems, TRS should be interpreted as the presence of a time-reversal-type antiunitary symmetry, rather than the physical time-reversal operation acting on real electron spins. These results naturally motivate the exploration of  {\zb  SU($N$) gauge fields with larger $N$} and their associated band topology in multicomponent cold-atom systems.

Spin-$3/2$ superconductivity provides another important motivation for studying multicomponent systems. In conventional superconductors, Cooper pairs are usually classified in terms of spin singlet and triplet channels formed by spin-$1/2$ quasiparticles~\cite{Sigrist1991RMP}. In contrast, spin-$3/2$ fermions allow a much richer pairing structure, including quintet and septet channels, and have attracted growing interest in the context of unconventional superconductivity~\cite{Ho1999,Yip1999,Brydon2016,Dutta2021,Wu2003,Wu2006,Yang2016,meinert2016,Roy2019,boettcher2016,boettcher2018,Venderbos2018,Jeong2021}. A prominent class of candidate materials is the half-Heusler superconductors, such as YPtBi and LuPtBi, where the low-energy quasiparticles can be described by effective $j=3/2$ states and unconventional pairing channels may 
arise~\cite{butch2011,bay2012,tafti2013,pan2013,nakajima2015,Kim2018,Radmanesh2018}. One particularly striking feature of multicomponent superconductors is the possible emergence of Bogoliubov Fermi surfaces~\cite{Agterberg2017,Brydon2018,menke2019,link2020,yuan2018,lapp2020,setty2020}, namely stable zero-energy quasiparticle surfaces in the superconducting state. Such gapless structures can appear in time-reversal-symmetry-breaking superconductors with nontrivial internal pairing structure and are protected by a {\zb $\mathbb{Z}_2$ Pfaffian} topological invariant \cite{Agterberg2017}. From this perspective, ultracold spin-$3/2$ atoms offer a complementary and highly controllable platform for exploring the interplay between high-spin pairing, synthetic spin-orbit coupling, and topological Bogoliubov quasiparticles.

In this work, we study topological band structures and interaction-induced superfluidity in square-lattice systems with homogeneous SU($N$) non-Abelian gauge fields. We first construct an explicit SU(4) gauge-field model related to the Hofstadter model with flux $\alpha=1/4${\zb ~\cite{Hofstadter1976}}. {\zb Among the four resulting bands, the lowest and highest are isolated, each carrying a Chern number $C=-4$, whereas the two middle bands touch at 16 Dirac points and jointly carry a Chern number $C=8$}. We then generalize the construction to SU($N$) systems and identify an even-odd structure in the band topology: {\zb when $N$ is odd, all bands are isolated and carry nonzero Chern numbers; 
 when $N$ is even, the two middle bands touch at $N^{2}$ Dirac points, while all other bands remain isolated and topologically nontrivial}. 
 In particular, the uppermost and lowermost bands become increasingly flat as $N$ increases, accompanied by a more uniform Berry curvature distribution, suggesting a possible route toward nearly ideal flat Chern bands with large Chern numbers. 
 {\zb We further focus on the SU(4) spin-$3/2$ model, exploring its superfluid phases in the presence of on-site attractive interactions and partial filling.} In the absence of the non-Abelian gauge field, the quintet pairing channels are degenerate due to the hidden SO(5) symmetry. We show that the SU(4) gauge field explicitly breaks this degeneracy and selects distinct nematic quintet pairing states depending on the chemical potential. When the chemical potential lies in the uppermost or lowermost nearly flat band, the quasiparticle spectrum is fully gapped. By contrast, in the middle-band regime, the selected quintet state can give rise to Bogoliubov Fermi surfaces protected by a {\zb $\mathbb{Z}_2$}  Pfaffian invariant.

The remainder of this paper is organized as follows. In Sec.\ref{II}, we introduce the theoretical model on a square lattice, construct an explicit form of an SU(4) non-Abelian gauge field, and demonstrate the emergence of topological band structures. In Sec.\ref{III}, we extend the construction to SU($N$) topological insulators and analyze the general features of their band structures and {\zb Berry curvature distributions}. In Sec.\ref{IV}, we focus on the superfluid phases of the SU(4) model. {\zb By varying the chemical potential, we identify both fully gapped nematic superfluid states and Bogoliubov Fermi surfaces. We discuss their respective properties and stability}. In Sec.\ref{V}, we discuss our results and conclude the paper.

\section{Model of SU(4) topological insulator}\label{II}
	
We consider a system of cold atoms hopping on a two-dimensional square lattice in the presence of a homogeneous non-Abelian gauge field. Utilizing the Peierls substitution, the tight-binding Hamiltonian in real space can be written as:
\begin{equation}
H=-t\sum_{i}(\psi_i^{\dagger}e^{-i{\zb A_x}}\psi_{i+\bm{\hat{x}}}+\psi_i^{\dagger}e^{-i{\zb A_y}}\psi_{i+\bm{\hat{y}}}+{\rm H.c.}){\zb ,}
\label{eq:tb ham}
\end{equation}
where $\psi_{i}=(c_{i,s},c_{i,s-1},...,c_{i,-s})^{T}$ is {\zb a} spinor operator, {\zb whose components $c_{i,s}$ denote
annihilation operators at site $i$ for the magnetic quantum number  $s$.  The parameter} $t$ is the nearest-neighbor hopping amplitude, and the gauge potentials {\zb $A_x$ and $A_y$} are matrix-valued operators that modify the hopping phases. {\zb Throughout this work, we set $t=1$ as the unit of energy. }

The link matrices appearing in the hopping terms $U_\mu=e^{-iA_\mu}$ are required to be elements of SU($N$). This choice defines the synthetic SU($N$) gauge-field model considered in this work. 
Physically, $U_\mu$ describes a unitary rotation among the $N$ internal atomic states during hopping. Similar matrix-valued hopping structures have been discussed in earlier proposals for non-Abelian optical lattices and cold-atom lattice gauge fields \cite{Osterloh2005,Goldman2009}.
The {\zb gauge potentials} $A_\mu$ are Hermitian traceless $N\times N$ matrices which can be expanded as
\begin{equation}
A_\mu=\sum_{a=1}^{N^2-1} A_\mu^a T^a,\qquad \mu=x,y,
\label{eq:gauge field}
\end{equation}
where $T^a$ are the $N^2-1$ generators of $\mathfrak{su}(N)$ {\zb(see details in Appendix~\ref{Appendixa})}. The gauge fields are non-Abelian when $[A_x,A_y]\neq 0$.

Starting from {\zb Eq.~\eqref{eq:tb ham}, we perform a Fourier transformation from the real space to} the momentum space and obtain the Bloch Hamiltonian
\begin{equation}
\mathcal{H}(\mathbf{k}) = -2t [\cos(k_x - A_x) + \cos(k_y - A_y)].
\end{equation}
Throughout the lattice constant is set to unity for notational simplicity.
Following the gauge-equivalence construction of Ref.~\cite{Barnett2012}, a Hofstadter model with Abelian flux $\alpha=1/N$ per plaquette can be mapped to a homogeneous SU($N$) non-Abelian gauge-field model by a position-dependent rotation in the internal space, we specialize this construction to $N=4$ in this section, corresponding to the flux $\alpha=1/4$. The two gauge potentials $A_x$ and $A_y$ read

\begin{equation*}
A_x=\frac{\pi}{4}
\begin{pmatrix}
    0&1-i &1 & 1+i\\
	1+i& 0& 1-i &1\\
	1&1+i &0 &1-i\\
	1-i&1 &1+i &0
\end{pmatrix},
\end{equation*}
\begin{equation}
A_y=\frac{\pi}{2}
\begin{pmatrix}
	\frac32 & & & \\
	& \frac12 & &\\
	& & -\frac12 &\\
	& & & -\frac32
\end{pmatrix}. 
\end{equation}
With these gauge potentials, {\zb the explicit matrix form of the four-band Bloch Hamiltonian is given by 
\begin{eqnarray}
\mathcal{H}(\mathbf{k}) = -2t [U_{x}\cos(k_x - \Lambda_x)U_{x}^{\dag} + \cos(k_y - A_y)],
\end{eqnarray}
where  $U_{x}$ is the unitary matrix that diagonalizes $A_{x}$,  and $\Lambda_{x}=\text{diag}\{3\pi/4,\pi/4,-\pi/4,-3\pi/4\}$ 
is the diagonalized form of $A_{x}$, satisfying $U_{x}\Lambda U_{x}^{\dag}=A_{x}$. }

\begin{figure}[t]
\centering
\includegraphics[width=0.5\textwidth]{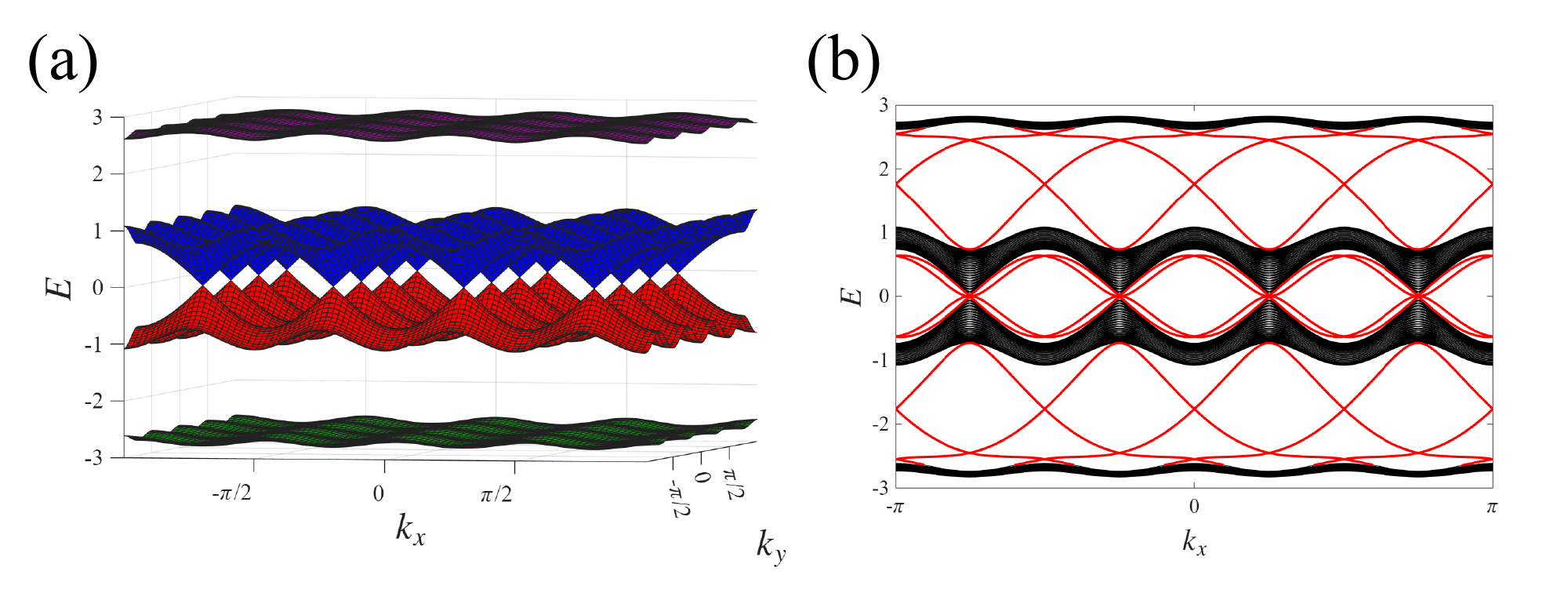}
\caption{(a) Bulk band structure of the SU(4) non-Abelian gauge-field model. The spectrum consists of four subbands. The lowest and highest bands are isolated, while the two middle bands touch at Dirac points. (b) Energy spectrum in a strip geometry. The red curves denote edge modes crossing the bulk gaps, reflecting the nonzero Chern numbers of the isolated bulk bands. }\label{fig1}
\end{figure}

Figure~\ref{fig1} shows the resulting band structure.
A characteristic feature of the SU(4) spectrum is the band touching between the two middle bands. In the full Brillouin zone, we find 16 Dirac points protected by the lattice and gauge field structure as shown in Fig.~\ref{fig1}(a). The lowest and highest bands are isolated from the other bands and carry nonzero Chern numbers. We find $C_1=C_4=-4$, where {\zb the subscript of $C_{m}$ refers to the band index and} the bands are ordered from bottom to top. {\zb Furthermore, neglecting the mutual contribution from the Dirac points yields $C_{2}=C_{3}=4$, in agreement with the constraint that the total Chern number summed over all bands vanishes.}  The nonzero Chern numbers of the isolated bands are further confirmed by the strip spectrum in Fig.~\ref{fig1}(b), where chiral in-gap edge modes appear inside the bulk gaps. Therefore, the SU(4) model realizes lattice quantum Hall states when the Fermi energy lies in the gaps above the lowest band or below the highest band.

We now briefly discuss the symmetries of the SU(4) Bloch Hamiltonian. 
The model does not preserve time-reversal symmetry. This is consistent with the nonzero Chern numbers of the isolated bands and with its interpretation as a lattice quantum Hall state. 
Nevertheless, the Hamiltonian has inversion symmetry. The inversion operation can be represented in the internal space by
\begin{equation}
\mathcal{P}=e^{-i\pi/4}
\begin{pmatrix}
0&0&0&1\\
0&0&-i&0\\
0&-1&0&0\\
i&0&0&0
\end{pmatrix},
\label{eq:inversion}
\end{equation}
under which
\begin{equation}
\mathcal{P} {\zb\mathcal{H}}(\mathbf{k}) \mathcal{P}^{-1}={\zb\mathcal{H}}(-\mathbf{k}).
\end{equation}
Equivalently, this follows from
\begin{equation}
\mathcal{P} A_x \mathcal{P}^{-1}=-A_x,\qquad
\mathcal{P} A_y \mathcal{P}^{-1}=-A_y.
\end{equation}
{\zb The inversion symmetry is important for our later discussion of the stability of Bogoliubov Fermi surfaces.

The spectrum shown in Fig.~\ref{fig1} is symmetric about 
$E=0$, which signals the presence of either chiral symmetry or subchiral 
symmetry~\cite{Mo2024SCS,Biao2024SCS,Liu2024SCS}. An analysis of the Hamiltonian confirms that it indeed possesses chiral symmetry, satisfying
\begin{eqnarray}
\mathcal{S}\mathcal{H}(\mathbf{k})\mathcal{S}^{-1}=-\mathcal{H}(\mathbf{k}), 
\end{eqnarray}
where the unitary and Hermitian matrix $\mathcal{S}$ takes the explicit form
\begin{equation}
\mathcal{S}=
\begin{pmatrix}
	 & & 1 & \\
	&  & & -1\\
	1 & &  &\\
	& -1 & & 
\end{pmatrix}. 
\end{equation}
The presence of chiral symmetry indicates that the Dirac points formed by the two middle bands are not accidental band 
crossings but are instead topological in origin. Each Dirac point is characterized by a 
winding number defined along a closed contour encircling it, taking a value of either 
$+1$ or $-1$, with the total sum over all Dirac points being zero.  
}

{\zb Before ending this section, it is worthwhile to compare the conventional gapless Dirac cones found in two-band models with 
$\mathcal{PT}$ or chiral symmetry---such as the tight-binding model for monolayer graphene~\cite{Neto2009graphene}---with those formed by the 
two middle bands in our system. In the conventional two-band case, the Berry curvature vanishes identically throughout 
the Brillouin zone except at the Dirac points, where it is ill-defined. In stark contrast, the two middle bands in our system exhibit finite Berry curvatures that yield a combined Chern number $C=8$. Consequently, despite being well separated from the highest and lowest bands, 
these two middle bands cannot be captured by any two-band tight-binding model, since the sum of Chern numbers over the two 
bands in any such model must necessarily be zero. In this sense, the Dirac cones formed by bands with nonzero Chern numbers 
are anomalous and are expected to give rise to physical phenomena distinct from those associated with conventional gapless Dirac cones.}

\section{GENERAL SU(N) BAND TOPOLOGY AND QUANTUM GEOMETRY}\label{III}
Having established the SU(4) model, we now extend the construction to a general SU($N$) non-Abelian gauge field. 
{\zb The explicit tight-binding Hamiltonian for the general SU($N$) model is provided in Appendix~\ref{Appendixb}.}
Instead of focusing on the explicit matrix form of the gauge potentials, we are interested here in the universal band-topological features that emerge when the number of internal components is increased. The SU(4) case discussed above already reveals two important properties:
the lowest and highest bands are isolated Chern bands, while the two middle bands touch at Dirac points. We show below that this structure is part of a more general even--odd pattern of SU($N$) topological {\zb insulators}.

\begin{figure}[t]
\centering
\includegraphics[width=0.5\textwidth]{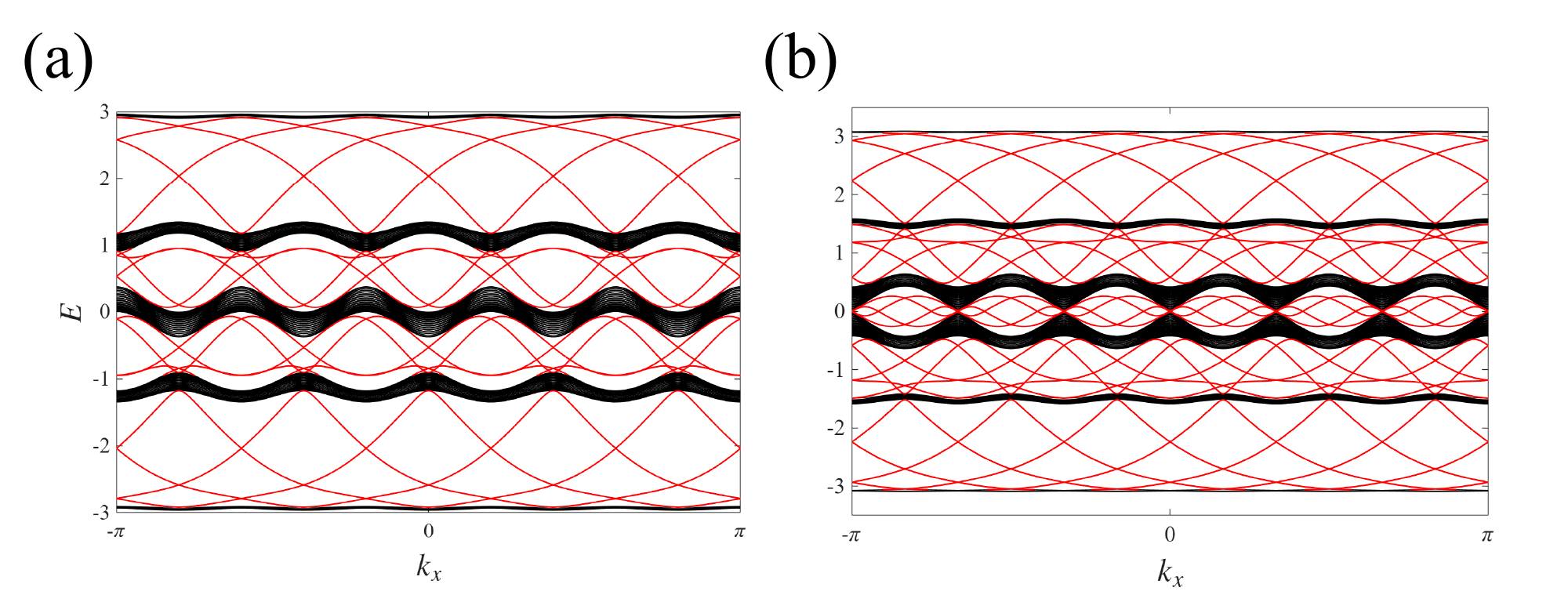}
\caption{
Energy spectra of the SU($N$) non-Abelian gauge-field models in a strip geometry with open boundary condition along the $y$ direction and periodic boundary condition along the $x$ direction. 
(a) Spectrum of the SU(5) model. The five bulk bands are separated by gaps, 
with Chern numbers $(-5,-5,20,-5,-5)$. 
(b) Spectrum of the SU(6) model. The two middle bands touch at Dirac points,
forming $36$ Dirac cones in the Brillouin zone, while the isolated bands carry
Chern number $-6$. Red curves denote chiral edge modes connecting bulk bands.
}\label{fig2}
\end{figure}

Figure~\ref{fig2} shows the spectra for the SU(5) and SU(6) models {\zb under a strip geometry}. For {\zb the} SU(5) {\zb case}, all five bands are separated by bulk gaps, and the Chern numbers are
\begin{equation}
(C_1,C_2,C_3,C_4,C_5)=(-5,-5,20,-5,-5).
\end{equation}
The middle band carries a large positive Chern number, while the remaining bands  {\zb each} carry {\zb an equal} Chern number. For the SU(6) case, the spectrum contains six bands. Similar to the SU(4) case, the two middle bands touch each other. In the full Brillouin zone, we find $36$ Dirac points. The band Chern numbers are
\begin{equation}
(C_1,C_2,C_3,C_4,C_5,C_6)=(-6,-6,12,12,-6,-6).
\end{equation}
Similarly, here the Chern numbers of the two middle bands should be understood after resolving the band-touching points, while the total Chern number carried by the middle two-band subspace is $24$.

From the results above, we identify a simple even--odd
structure. For odd $N$, all bands are isolated. The middle band carries Chern
number
\begin{equation}
C_{\rm mid}=N(N-1),
\end{equation}
whereas the other $N-1$ bands carry
\begin{equation}
C=-N.
\end{equation}
For even $N$, the two middle bands touch at Dirac points. The number of Dirac cones is
\begin{equation}
N_D=N^2.
\end{equation}
The remaining $N-2$ isolated bands {\zb each} carry a Chern number $-N$.
Besides the Chern-number pattern discussed above, another important feature of the SU($N$) models is the progressive flattening of the uppermost and lowermost bands as $N$ increases.
Therefore, increasing the number of internal components provides a natural way to realize {\zb well-separated}, nearly flat Chern bands with large topological indices.

\begin{figure}[t]
\centering
\includegraphics[width=0.45\textwidth]{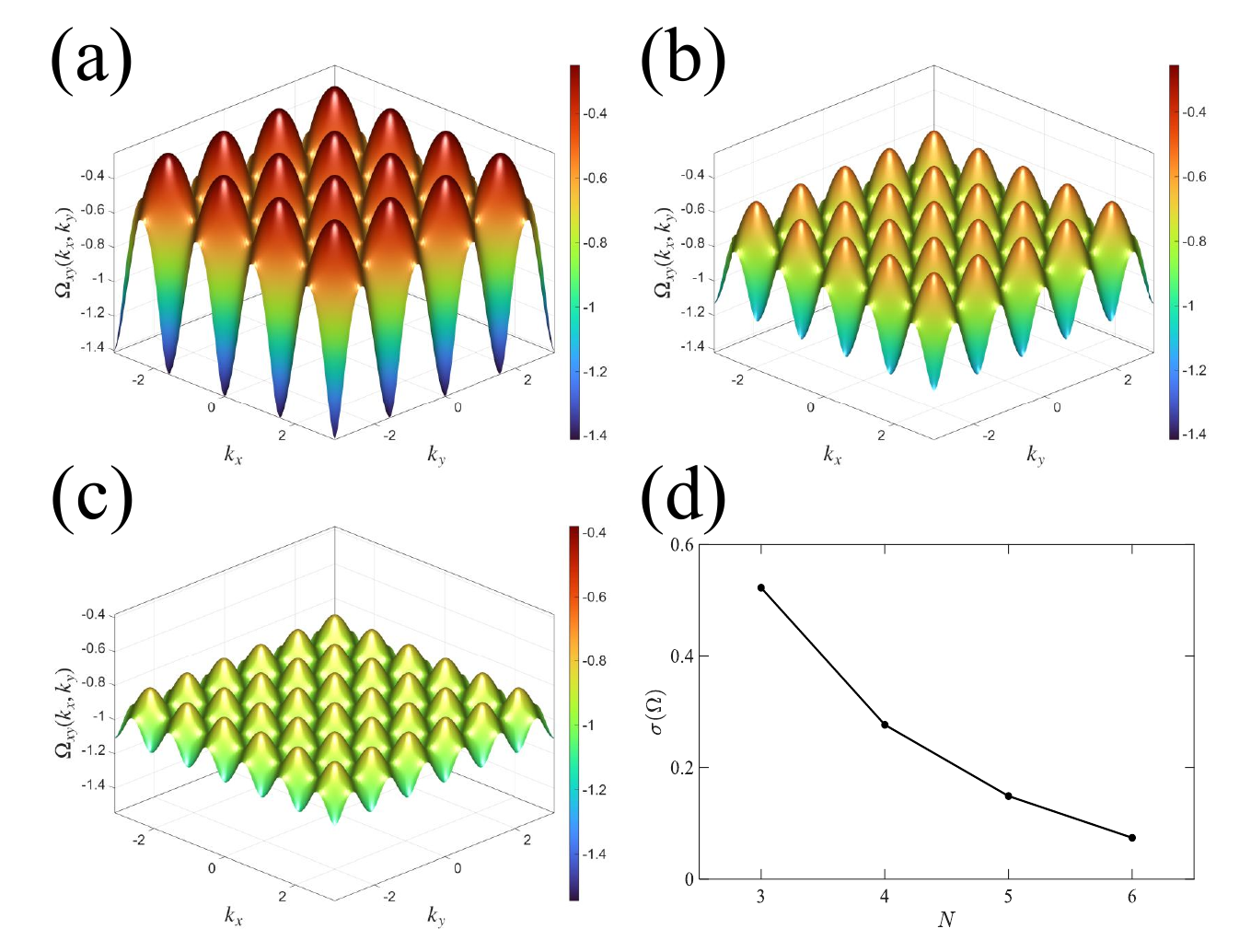}
\caption{
Berry curvature distribution of the lowermost Chern band in the SU($N$) non-Abelian gauge-field models.
(a)--(c) Berry curvature $\Omega_{xy}(\mathbf{k})$ for the SU(4), SU(5), and SU(6) models, respectively. As $N$ increases, the Berry curvature becomes progressively more uniform over the Brillouin zone.
(d) Standard deviation $\sigma(\Omega)$ of the Berry curvature for the {\zb lowermost, equivalently uppermost}, Chern band as a function of $N$. 
}\label{fig3}
\end{figure}

To further characterize the quality of these nearly flat Chern bands, we examine the Berry curvature distribution of the lowermost band {\zb in the Brillouin zone. }
Figures \ref{fig3}(a)--(c) show the Berry curvature $\Omega_{xy}(\mathbf{k})$ for the SU(4), SU(5), and SU(6) models, respectively. As $N$ increases, the spatial variation of $\Omega_{xy}(\mathbf{k})$ over the Brillouin zone is strongly suppressed, {\zb and} the Berry curvature becomes much more uniformly distributed.
This tendency is quantified in Fig.~\ref{fig3}(d), where we plot the standard deviation $\sigma(\Omega)$ of the Berry curvature for the {\zb lowermost, equivalently uppermost}, Chern band. The monotonic decrease of $\sigma(\Omega)$ with increasing $N$ demonstrates that the Berry curvature
approaches a uniform distribution in the large-$N$ regime. Together with the suppressed band dispersion, this result indicates that the uppermost and lowermost SU($N$) bands gradually approach ideal flat Chern bands.

Flat Chern bands are of special interest because they provide a lattice analogue of Landau levels and may support strongly correlated topological phases 
when {\zb strong repulsive interactions} are included~\cite{Tang2011,Sun2011,Neupert2011,sheng2011,regnault2011,yang2012,liu2012,bergholtz2013}, such as the fractional Chern insulators that have recently been observed in various moir\'{e} systems~\cite{Cai2023,Zeng2023,Park2023}. 
Besides a narrow bandwidth and a finite band gap, the distribution of Berry curvature and quantum metric over the Brillouin zone plays an important role in determining how closely a Chern band resembles an ideal Landau level~\cite{parameswaran2013,roy2014,jackson2015,Ledwith2023}.  Although the emergence of fractional Chern insulator phases also depends on the interaction form and filling fraction, the single-particle properties found here are favorable for their realization. Notably, the uppermost and lowermost SU($N$) bands combine a large Chern number, weak dispersion, and an increasingly uniform Berry curvature, {\zb making them promising platforms for more exotic unconventional fractional Chern insulators without Landau-level analogs}. 

{\zb In cold-atom systems, both the strength and sign (repulsive or attractive) of interatomic interactions 
can be tuned over a wide range, for instance via Feshbach resonances. Rather than exploring strongly 
correlated phases in the strongly repulsive regime, here we focus on attractive interactions 
and investigate the resulting superfluid phases within the mean-field framework. As an initial study, we consider an on-site  attractive 
Hubbard interaction and examine superfluidity in the SU(4) model across various filling fractions.}


\section{superfluidity in the SU(4) model}\label{IV}
In the ultracold regime, short-range interactions are dominated by the $s$-wave scattering channel. Therefore, we restrict ourselves to local onsite {\zb attractive Hubbard} interactions, which {\zb give rise to} $s$-wave (i.e., zero orbital angular momentum) pairing in the lattice model. Constrained by the Pauli exclusion principle, we consider only two independent interaction channels with total spins $J=0$ and $J=2$. The {\zb full spin-$3/2$} Hamiltonian reads
\begin{align}
H&=-\sum_{\braket{i,j},\alpha\beta}{\zb t_{\alpha\beta}c_{i\alpha}^{\dagger}c_{j\beta}}-\mu\sum_{i\alpha}c_{i\alpha}^{\dagger}c_{i\alpha}\nonumber\\&-U_0\sum_{i}P_{0,0}^{\dagger}(i)P_{0,0}(i)-U_2\sum_{i,m}P_{2,m}^{\dagger}(i)P_{2,m}(i),
\label{eq:interaction}
\end{align}
where $\alpha,\beta$ label the four spin components. {\zb The gauge potentials are absorbed implicitly into the hopping amplitudes $t_{\alpha\beta}$, 
and $\mu$ denotes the chemical potential.}  The onsite pairing operators are defined as $P^\dagger_{J,m}(i)=\sum_{\alpha\beta}\braket{\frac32,\frac32;J,m|\frac32,\frac32;\alpha,\beta}c^\dagger_{i,\alpha}c^\dagger_{i,\beta}$. The explicit coefficients in $P^\dagger_{J,m}$ can be read off from {\zb Appendix \ref{Appendixc}}. With the minus signs in Eq.~(\ref{eq:interaction}), positive \(U_0\) and \(U_2\)
correspond to attractive interactions in the singlet and quintet channels, respectively. 

Within the {\zb uniform-pairing} ansatz, we introduce the singlet and quintet order parameters as
\begin{equation}
\Delta_{J,m}=\frac{U_J}{N_s}\sum_i \braket{P_{J,m}(i)}=U_J\braket{P_{J,m}(i)}{\zb ,}
\end{equation}
where the order parameters are assumed to be site-independent. With this convention, the interaction part is {\zb decomposed} as
\begin{align}
H_{\rm int}^{\rm MF}
=&-\sum_i\left[
\Delta_{0,0}P_{0,0}^\dagger(i)
+\Delta_{0,0}^*P_{0,0}(i)
\right]\nonumber \\
&-\sum_{i,m}\left[
\Delta_{2,m}P_{2,m}^\dagger(i)
+\Delta_{2,m}^*P_{2,m}(i)
\right]\nonumber \\
&+N_s\left(
\frac{|\Delta_{0,0}|^2}{U_0}
+\sum_{m=-2}^{2}\frac{|\Delta_{2,m}|^2}{U_2}
\right),
\end{align}
where $N_s$ is the number of lattice sites. After the Fourier transformation
\begin{equation}
c_{i\alpha}=
\frac{1}{\sqrt{N_s}}
\sum_{\mathbf{k}}
e^{i\mathbf{k}\cdot\mathbf{r}_i}
c_{\mathbf{k}\alpha},
\end{equation}
the onsite pairing operators create Cooper pairs with zero center-of-mass momentum,
\begin{equation}
\sum_i P_{J,m}^{\dagger}(i)
=
\sum_{\mathbf{k},\alpha\beta}
C_{\alpha\beta}^{Jm}
c_{\mathbf{k}\alpha}^{\dagger}
c_{-\mathbf{k}\beta}^{\dagger},
\end{equation}
where \(C_{\alpha\beta}^{Jm}\) denotes the corresponding Clebsch--Gordan coefficient. Equivalently, the self-consistency equations can be evaluated from the anomalous correlation functions as
\begin{equation}
\Delta_{J,m}
=
\frac{U_J}{N_s}
\sum_{\mathbf{k},\alpha\beta}
\left(C_{\alpha\beta}^{Jm}\right)^*
\left\langle
c_{-\mathbf{k}\beta}
c_{\mathbf{k}\alpha}
\right\rangle .
\end{equation}
Therefore, the onsite pairing order parameters are momentum independent in the original spin basis. The momentum dependence of the 
{\zb Bogoliubov-de Gennes (BdG)} Hamiltonian comes from the normal-state Bloch Hamiltonian
\(H_t(\mathbf{k})\), while the pairing matrix \(\hat{\Delta}\) is a \(k\)-independent antisymmetric matrix.

In the spinor basis
\(
\psi_{\mathbf{k}}
=
(
c_{\mathbf{k},3/2},
c_{\mathbf{k},1/2},
c_{\mathbf{k},-1/2},
\allowbreak c_{\mathbf{k},-3/2}
)^T,
\)
we introduce the Nambu spinor
$
\Psi_{\mathbf{k}}
=
\left(
\psi_{\mathbf{k}},
\psi_{-\mathbf{k}}^{\dagger}
\right)^T .
$
The mean-field Hamiltonian can then be written as
$
H_{\rm MF}
=
\frac{1}{2}
\sum_{\mathbf{k}}
\Psi_{\mathbf{k}}^{\dagger}
\mathcal{H}_{\rm BdG}(\mathbf{k})
\Psi_{\mathbf{k}}
+{\rm const.},
$
where

\begin{equation}
\mathcal{H}_{\rm BdG}(\mathbf{k})
=
\begin{pmatrix}
{\zb \mathcal{H}_t}(\mathbf{k})-\mu
&
\hat{\Delta}
\\[4pt]
\hat{\Delta}^{\dagger}
&
-\mathcal{H}_t^T(-\mathbf{k})+\mu
\end{pmatrix}.
\end{equation}
Here \(\mathcal{H}_t(\mathbf{k})\) denotes the normal-state Bloch Hamiltonian. To write the pairing matrix in a compact form, we define
$\Delta_{\pm}
\equiv
(\Delta_{2,0}\pm \Delta_{0,0})/\sqrt{2}
$.
Then the pairing matrix is given by

\begin{equation}
\hat{\Delta}
=
\begin{pmatrix}
0 & \Delta_{2,2} & \Delta_{2,1} & \Delta_{+} \\
-\Delta_{2,2} & 0 & \Delta_{-} & -\Delta_{2,-1} \\
-\Delta_{2,1} & -\Delta_{-} & 0 & -\Delta_{2,-2} \\
-\Delta_{+} & \Delta_{2,-1} & \Delta_{2,-2} & 0
\end{pmatrix}.
\end{equation}

\subsection{Symmetry considerations}

Before presenting the self-consistent mean-field results, we first discuss the symmetry of the interacting spin-$3/2$ model. In the absence of the non-Abelian gauge field, namely for $A_x=A_y=0$, the hopping term is diagonal in the internal spin space,
\begin{equation}
H_t=-t\sum_{\langle i,j\rangle,\alpha}{\zb c_{i\alpha}^{\dagger}c_{j\alpha}}.
\end{equation}
This kinetic term is invariant under a global SU(4) rotation among the four spin components
\(\psi_i=(c_{i,3/2},c_{i,1/2},c_{i,-1/2},c_{i,-3/2})^T\). Following Ref.~\cite{Wu2003}, these two pairing channels can be reorganized into an SO(5) scalar and an SO(5) five-vector. Therefore, for generic values of $U_0$ and $U_2$, the onsite interaction is invariant under the hidden SO(5) symmetry. Equivalently, the interaction breaks the global SU(4) symmetry of the kinetic term down to SO(5). A higher symmetry emerges on the line $U_0=U_2$. In this case, the singlet and quintet pairing channels become degenerate and together 
form the six-dimensional antisymmetric tensor representation of SU(4){\zb~\cite{Majid2023}}. 
The full gauge-free Hamiltonian is {\zb accordingly} promoted from the generic SO(5) symmetry to a global SU(4) symmetry.

We now turn to the model with the homogeneous non-Abelian gauge field introduced in {\zb Sec.~\ref{II}}. The hopping term becomes
\begin{equation}
H_t=-t\sum_{i,\mu=x,y}
\left(\psi_i^\dagger e^{-iA_\mu}\psi_{i+\hat{\mu}}+{\rm H.c.}\right).
\end{equation}
Here $e^{-iA_\mu}$ are fixed {\zb SU}(4) link matrices acting on the internal spin space. The hopping term is invariant only if
\begin{equation}
G^\dagger e^{-iA_\mu}G=e^{-iA_\mu},\qquad \mu=x,y .
\end{equation}
where \(G\) is a global internal rotation . For the SU(4) gauge field considered in {\zb Sec.~\ref{II}}, 
this condition is not satisfied by the full SO(5) group. Hence the non-Abelian gauge field acts as 
a synthetic spin-orbit coupling and explicitly breaks the hidden SO(5) symmetry.

\subsection{Pairing phase diagram}
\begin{figure}[t]
\centering
\includegraphics[width=0.52\textwidth]{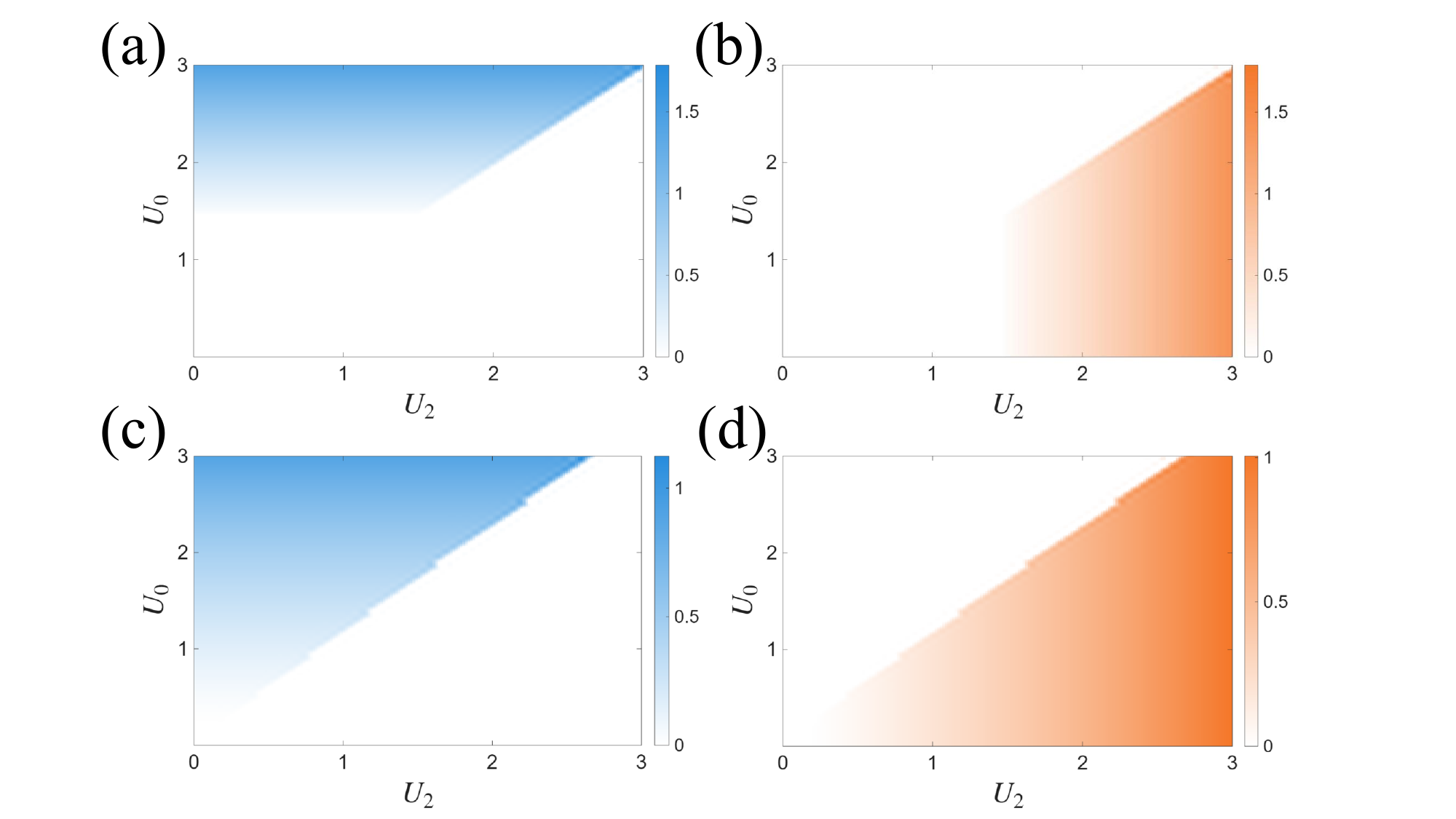}
\caption{
Self-consistent pairing amplitudes of the spin-$3/2$ attractive Hubbard model.
(a,b) Results for the gauge-free model with $A_x=A_y=0$ at $\mu=-1$.
The blue color scale in (a) denotes the singlet pairing strength $|\Delta_0|$, while the orange color scale in (b) denotes the quintet pairing strength
$\Delta_2=\left(\sum_{m=-2}^{2}|\Delta_{2,m}|^2\right)^{1/2}$.
(c,d) Corresponding results for the SU(4) non-Abelian gauge-field model at $\mu=-2.75$, where the chemical potential lies in the lowest nearly flat band of the SU(4) spectrum. The meanings of the color scales in (c) and (d) are the same as those in (a) and (b), respectively.
}
\label{fig4}
\end{figure}

We now solve the mean-field {\zb self-consistent} equations for both the gauge-free model and the SU(4) non-Abelian gauge-field model. The resulting pairing phase diagrams are shown in Fig.~\ref{fig4}. To characterize the quintet pairing sector, we define the total quintet pairing amplitude as
\begin{equation}
\Delta_2=\left(\sum_{m=-2}^{2}|\Delta_{2,m}|^2\right)^{1/2}.
\end{equation}
In the absence of the non-Abelian gauge field, the phase diagram shows a simple structure. The singlet and quintet pairing phases are separated by the line $U_0=U_2$. With our convention that $U_0$ and $U_2$ denote the attractive interaction strengths in the singlet and quintet channels, respectively. {\zb The} singlet pairing state is energetically favored for $U_0>U_2$, whereas the quintet pairing state is favored for $U_2>U_0$ {\zb [see Figs.~\ref{fig4}(a) and \ref{fig4}(b)]}. This behavior is consistent with the symmetry analysis above: without the gauge field, the five quintet components form an SO(5) vector, and the line $U_0=U_2$ corresponds to the enhanced SU(4)-symmetric point where the singlet and quintet channels become degenerate.

The situation changes qualitatively once the SU(4) non-Abelian gauge field is introduced. The pairing phase diagram is no longer determined solely by the relative magnitude of $U_0$ and $U_2$. In particular, for $\mu=-2.75$, where the chemical potential lies in the lowest nearly flat band of the SU(4) spectrum, the self-consistent solution on the line $U_0=U_2$ is found to be a quintet pairing state 
rather than a degenerate singlet--quintet state {\zb [see Figs.~\ref{fig4}(c) and \ref{fig4}(d)]}. This result indicates that the non-Abelian gauge field 
not only modifies the normal-state band structure, but also selects the energetically favorable pairing channel in the superconducting state.

We further examine the structure of the quintet pairing order parameter. When $A_x=A_y=0$, the five quintet pairing channels are degenerate. The symmetry has a direct implication for the quintet pairing order parameters. The five components of the quintet order parameter, \(\vec{\Delta}_2=(\Delta_{2,2},\Delta_{2,1},\Delta_{2,0},
\allowbreak \Delta_{2,-1},\Delta_{2,-2})\)
form a five-dimensional vector representation of SO(5). Consequently, different self-consistent solutions with the same value of $\Delta_2$ but different relative weights of $\Delta_{2,m}$ are connected by SO(5) rotations and are therefore energetically degenerate. This explains why, in the gauge-free calculation, the quintet phase is characterized by a finite total amplitude $\Delta_2$, while no unique quintet component is selected.
\begin{figure}[t]
\centering
\includegraphics[width=0.45\textwidth]{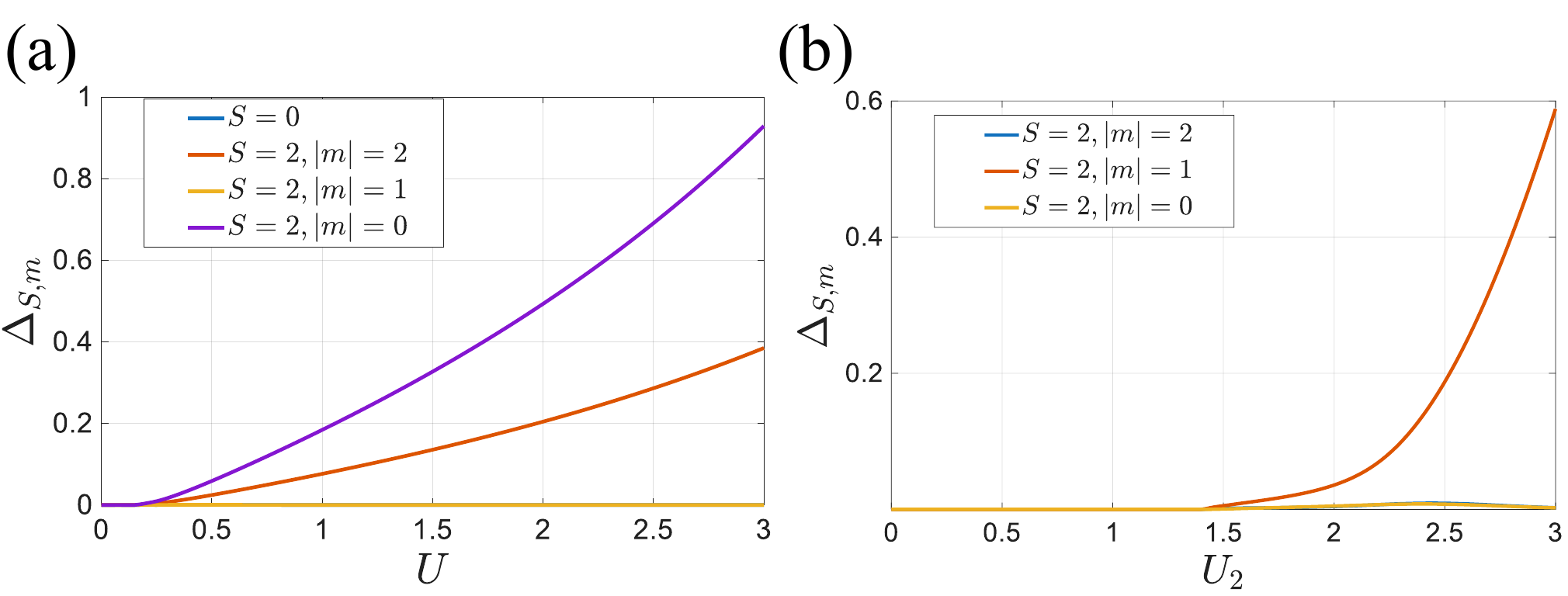}
\caption{
Self-consistent pairing amplitudes in the SU(4) gauge-field model.
(a) Results for $\mu=-2.75$ with the chemical potential in the lowest nearly flat band, obtained along the line $U_0=U_2=U$. The singlet and all quintet pairing components are plotted as functions of $U$.
(b) Results for $\mu=-0.3$ with the chemical potential below the Dirac points, obtained along $U_0=0$. The five quintet components $\Delta_{2,m}$ are plotted as functions of $U_2$.
}
\label{fig5}
\end{figure}

We next focus on the internal structure of the quintet order parameter in the SU(4) gauge-field model. As discussed above, the non-Abelian gauge field explicitly breaks the hidden SO(5) symmetry. As a result, the five quintet pairing channels are no longer required to be degenerate. The gauge field can select particular linear combinations of the quintet components, leading to the distinct self-consistent pairing structures discussed below.

This symmetry breaking is clearly reflected in Fig.~\ref{fig5}. For $\mu=-2.75$, where the chemical potential lies in the lowest nearly flat band, the self-consistent solution along $U_0=U_2=U$ is dominated by the $m=0$ and $m=\pm 2$ quintet components, while the $|m|=1$ components vanish. After fixing the overall $U(1)$ phase of the superconducting order parameter, the two components $\Delta_{2,2}$ and $\Delta_{2,-2}$ have a relative phase difference of $\pi$, whereas the $m=0$ component and the $|m|=2$ components have a relative phase difference of $\pi/2$. Therefore, the quintet order parameter takes the form
\begin{equation}
\Delta_2
\sim
|2,0\rangle
+iC\left(|2,2\rangle-|2,-2\rangle\right),
\end{equation}
where $C$ is a real coefficient determined by the self-consistent solution.

For $\mu=-0.3$, where the chemical potential lies below the Dirac points of the middle bands, the selected quintet structure is qualitatively different. In this case, the $m=0$ and $|m|=2$ components vanish, while only the $m=\pm1$ components remain finite. The relative phase between $\Delta_{2,1}$ and $\Delta_{2,-1}$ is $\pi/2$. Thus the quintet order parameter can be written as
\begin{equation}
\Delta_2
\sim
|2,1\rangle-i|2,-1\rangle ,
\end{equation}
up to an overall superconducting phase. These results show that the SU(4) non-Abelian gauge field not only favors the quintet pairing channel in certain parameter regimes, but also fixes the direction of the quintet vector in the five-dimensional order-parameter space. The selected pairing structure depends sensitively on the position of the chemical potential in the SU(4) band structure. When the chemical potential lies in the uppermost or lowermost nearly flat band, the favored quintet state involves the $m=0$ and $m=\pm2$ components. By contrast, in the middle-band region, the favored state is formed by the $m=\pm1$ components.

It should be emphasized that the above quintet order parameters should be understood as nematic quintet pairings rather than chiral $d+id$ states. Although relative phase factors appear in the spherical basis $|2,m\rangle$, this basis is intrinsically complex. After transforming to the real tensor basis of the spin-quintet representation, the selected order parameters can be chosen real up to an overall superconducting $U(1)$ phase. In this sense, the two self-consistent solutions correspond to nematic $d$-wave-like order parameters.

{\zb Comparing Fig~\ref{fig5}(a) with Fig.~\ref{fig5}(b),  we further find two qualitatively distinct features in the evolution of the pairing amplitude with interaction strength. First, when the chemical potential lies in the nearly flat Chern band, a notable pairing amplitude emerges at a relatively low interaction strength. This behavior is expected, since the density of states in a flat band is generally much higher than that in a dispersive band at a generic energy. Second, in the weak-pairing regime (e.g., defined by $|\Delta|<0.2$), the pairing amplitude increases almost linearly with interaction strength for the flat-band case, in contrast to the exponential growth observed when the chemical potential lies in the dispersive band. This linear scaling is a hallmark of flat-band superconductivity and is particularly noteworthy because it deviates from the conventional exponential dependence predicted by Bardeen–Cooper–Schrieffer theory, suggesting that flat bands may offer a promising route to high-temperature superconductivity~\cite{Kopnin2011,Kopnin2013}.}

\subsection{Bogoliubov Fermi surfaces}

We next examine the quasiparticle spectrum of the superconducting state. 
For the quintet states selected by the SU(4) gauge field, the excitation spectrum depends sensitively on the position of the chemical potential. 
When the chemical potential lies in the uppermost or lowermost nearly flat band, the self-consistent quintet order parameter mainly involves the $m=0$ and $m=\pm 2$ components. In this case, the Bogoliubov-de Gennes spectrum is fully gapped. By contrast, a qualitatively different behavior appears when the chemical potential lies in the middle-band region.
\begin{figure}[t]
\centering
\includegraphics[width=0.5\textwidth]{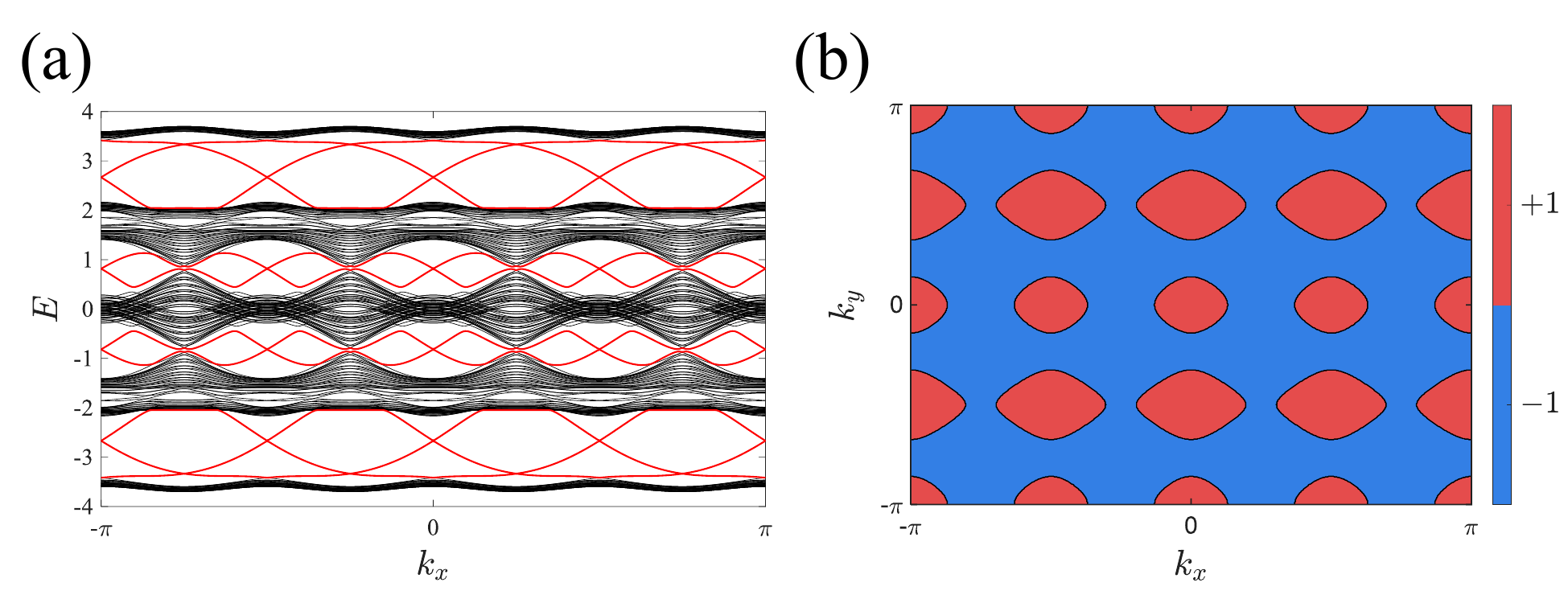}
\caption{
Bogoliubov Fermi surfaces in the SU(4) gauge-field superfluid when the chemical potential lies in the middle-band region.
The parameters are chosen as $\mu=-0.8$, $U_0=2$, and $U_2=3$, for which the self-consistent solution is dominated by quintet pairing.
(a) Bogoliubov-de Gennes quasiparticle spectrum. The zero-energy contours indicate the emergence of Bogoliubov Fermi surfaces.
(b) Sign of the Pfaffian topological invariant in the Brillouin zone. The red and blue regions denote positive and negative Pfaffian signs, respectively. The sign-changing boundaries coincide with the Bogoliubov Fermi surfaces shown in (a).
}
\label{fig6}
\end{figure}
Figure~\ref{fig6} shows the BdG spectrum for $\mu=-0.8$, $U_0=2$ and $U_2=3$. 
For this choice of parameters, the self-consistent solution is dominated by the $|m|=1$ quintet pairing channel $\Delta_2\sim|2,1\rangle-i|2,-1\rangle $. 
As shown in Fig.~\ref{fig6}(a), the quasiparticle spectrum is no longer fully gapped. Instead, the zero-energy excitations form closed contours in momentum space. These zero-energy contours are Bogoliubov Fermi surfaces{\zb~\cite{Agterberg2017}}, which appear only in the middle-band regime where the SU(4) gauge field selects the $|m|=1$ quintet pairing structure in our numerical results.

The stability of these Bogoliubov Fermi surfaces can be characterized by a {\zb$\mathbb{Z}_{2}$} Pfaffian topological invariant{\zb~\cite{Agterberg2017,Mo2025}}. 
Although the BdG Hamiltonian breaks time-reversal symmetry, it preserves particle-hole symmetry which takes the form 
$\mathcal{C} = \tau_x \mathcal{K}$, where \(\tau_x\) acts in the Nambu particle-hole space and $\mathcal{K}$ denotes complex conjugation.
In addition, {\zb as analyzed in Sec.~\ref{II}}, the normal-state Hamiltonian has inversion symmetry. In the BdG basis, given that the pairing matrix 
is {\zb even under inversion}, the corresponding inversion operator is
\begin{equation}
    \widetilde{\mathcal{P}}
    =
    \begin{pmatrix}
        \mathcal{P} & 0 \\
        0 & \mathcal{P}^{*}
    \end{pmatrix}{\zb,}
\end{equation}
where $\mathcal{P}$ is given by {\zb Eq.}~\ref{eq:inversion}. The combination of particle-hole symmetry and inversion symmetry gives an antiunitary symmetry $\mathcal{C}\widetilde{\mathcal{P}} = U_{\mathcal{C}\mathcal{P}} \mathcal{K}$ acting locally at each momentum, where \(U_{\mathcal{C}\mathcal{P}}\) is the unitary part of the \(\mathcal{C}\widetilde{\mathcal{P}}\) operation. 
This symmetry imposes
\begin{equation}
    U_{\mathcal{C}\mathcal{P}}
    \mathcal{H}_{\mathrm{BdG}}^{*}(\mathbf{k})
    U_{\mathcal{C}\mathcal{P}}^{\dagger}
    =
    -\mathcal{H}_{\mathrm{BdG}}(\mathbf{k}) .
\end{equation}
Therefore, one can define
\begin{equation}
    \widetilde{\mathcal{H}}(\mathbf{k})
    =
    \mathcal{H}_{\mathrm{BdG}}(\mathbf{k})
    U_{\mathcal{C}\mathcal{P}} .
\end{equation}
In the basis used here, \(U_{\mathcal{C}\mathcal{P}}^{T}=U_{\mathcal{C}\mathcal{P}}\). 
Using the Hermiticity of the BdG Hamiltonian and the above \(\mathcal{C}\widetilde{\mathcal{P}}\) constraint, we obtain
\begin{equation}
    \widetilde{\mathcal{H}}^{T}(\mathbf{k})
    =
    -\widetilde{\mathcal{H}}(\mathbf{k}) .
\end{equation}
Thus, \(\widetilde{\mathcal{H}}(\mathbf{k})\) is an antisymmetric matrix, and its Pfaffian is well defined. 
One can define the momentum-resolved Pfaffian invariant as
\begin{equation}
    \nu(\mathbf{k})
    =
    \mathrm{sgn}
    \left\{
    \mathrm{Pf}
    \left[
    \widetilde{\mathcal{H}}(\mathbf{k})
    \right]
    \right\}.
\end{equation}
The sign of this Pfaffian can change only when the BdG spectrum becomes gapless, i.e., when
$\det [ \mathcal{H}_{\mathrm{BdG}}(\mathbf{k}) ] = 0$. Thus, the boundary between regions with opposite Pfaffian signs corresponds precisely to the Bogoliubov Fermi surface.

This is confirmed in Fig.~\ref{fig6}(b), where we plot the $\nu(\mathbf{k})$ over the Brillouin zone. 
The red and blue regions represent positive and negative Pfaffian signs, respectively. The sign changing lines coincide with the zero-energy contours in the BdG spectrum. This agreement demonstrates that the Bogoliubov Fermi surfaces are not accidental gapless features, but are protected by the Pfaffian topology of the BdG Hamiltonian.

{\zb From Fig.~\ref{fig6}(b), it is readily seen that the Bogoliubov Fermi surfaces possess only 
$C_{2z}$ rotation symmetry. Since the Fermi surface of the normal-state Hamiltonian has 
$C_{4z}$ rotation symmetry, this reduction indicates that the quintet pairing lowers the rotational symmetry from 
$C_{4z}$ to $C_{2z}$. This is consistent with our previous analysis that, under this parameter condition, 
the resulting quintet pairing is nematic in nature. }

\section{discussions and conclusions}\label{V}

In this work, we have studied topological band structures and interaction-induced superfluid phases in square-lattice systems with homogeneous SU($N$) non-Abelian gauge fields. Starting from an SU(4) gauge-field model, we showed that the non-Abelian hopping matrices {\zb naturally} generate a topological band structure. {\zb Interestingly,  the number of spin components directly determines the key features of the band structure. 
Specially, we find that the lowest and highest bands each carry a Chern number $C=-4$, while the two middle bands touch at 16 
Dirac points and possesses a combined Chern number $C=8$. Extending the analysis to larger $N$, we identify an even--odd structure 
in the band topology.}  For odd $N$, all bands are separated by bulk gaps, with the middle band carrying a large Chern number $C_{\rm mid}=N(N-1)$, while {\zb each of the remaining bands has} $C=-N$. For even $N$, the two middle bands touch at $N^{2}$ Dirac points, whereas the other isolated bands {\zb each} carry  $C=-N$. Another important feature is that the uppermost and lowermost bands become increasingly flat as $N$ grows. At the same time, the Berry curvature of these bands becomes more uniformly distributed over the Brillouin zone. Therefore, increasing the number of internal components provides an efficient way to approach nearly ideal flat Chern bands with large Chern numbers. These results suggest that SU($N$) non-Abelian gauge-field systems may serve as a useful platform for exploring strongly correlated topological phases, including possible fractional Chern insulating states {\zb without Landau-level analogs}.

{\zb We further specialized to the spin-$3/2$ SU(4) model and studied the superfluid ground states
induced by on-site attractive interactions at partial filling.} In the absence of the non-Abelian gauge field, the spin-3/2 interaction channels possess a hidden SO(5) symmetry, under which the five quintet pairing components form a vector representation. As a consequence, different quintet pairing configurations with the same total amplitude are related by SO(5) rotations and are degenerate. Once the SU(4) non-Abelian gauge field is added, this degeneracy is lifted. The gauge field acts as a synthetic spin-orbit coupling and selects particular quintet pairing structures. When the chemical potential lies in the upper or lower nearly flat Chern band, the favored state is a fully gapped singlet or nematic quintet superfluid involving the $m=0$ and $m=\pm 2$ components. In contrast, when the chemical potential lies in the middle-band region, the self-consistent solution is dominated by either singlet or the $m=\pm 1$ quintet components.  {\zb In the quintet-dominated regime, the quasiparticle spectrum exhibits Bogoliubov Fermi surfaces, protected by the $\mathbb{Z}_{2}$ Pfaffian invariant.}

The model we discussed is {\zb directly} relevant to ultracold atomic systems. In contrast to solid-state materials, where spin-orbit coupling is constrained by crystal symmetry and by the properties of real electron spin, cold atoms allow the internal atomic states to be engineered synthetically. The {\zb $N$} internal components can be chosen from hyperfine states or other long-lived atomic states, and laser-assisted tunneling or Raman-coupling techniques can generate matrix-valued hopping amplitudes between them. In particular, the $N$-pod scheme provide a promising way to realize static non-Abelian gauge fields~\cite{Ruseckas2005,Burrello2011,Hasan2022,Juzeliunas2010}, i.e., SU($N$) spin-orbit coupling in optical lattices.  By designing the laser configuration in such schemes, one may engineer effective SU($N$) gauge potentials similar to those considered in this work. 

Looking ahead, {\zb there are two natural directions for extension. First, one may generalize the on-site 
attractive interactions to include nearest-neighbor couplings. As the interaction range increases, 
pairings with odd total spin angular momentum, as well as complex pairings with nonzero orbital angular 
momentum, become possible, potentially giving rise to a variety of topological superconducting phases.} 
{\zb Second, it would be valuable to explore the repulsive regime and investigate the exotic correlated phases 
that may emerge there. In summary,} our results suggest that synthetic 
SU($N$) gauge fields offer a versatile foundation for realizing topological bands {\zb with distinctive properties, 
thereby providing a flexible platform for exploring novel physical phenomena and unconventional interaction-driven phases.}

\section*{Acknowledgements}

W.Z. and Z.Y. {\zb acknowledge support from the Fundamental and Interdisciplinary Disciplines Breakthrough Plan of the Ministry of Education of China (Grant No. JYB2025XDXM403) and the Guangdong Basic and Applied Basic Research Foundation (Grant No. 2023B1515040023).}

\appendix

\section{$\mathfrak{su}$(4) algebra\label{Appendixa}}
In the main text, we mentioned that the non-Abelian gauge potential can be expanded in terms of the generators of the Lie algebra, as 
presented in Eq.~\ref{eq:gauge field}, In this section, we explicitly give one convenient basis of the $\mathfrak{su}(4)$ algebra, and these generators are denoted by $\lambda_a$ in the following.  

The fifteen generators are chosen as

\begin{equation*}
\lambda_1=
\begin{pmatrix}
    0 &1 &0 &0\\
    1 &0 &0 &0\\
    0 &0 &0 &0\\
    0 &0 &0 &0
\end{pmatrix},
\lambda_2=
\begin{pmatrix}
    0 &-i &0 &0\\
    i &0 &0 &0\\
    0 &0 &0 &0\\
    0 &0 &0 &0
\end{pmatrix},
\lambda_4=
\begin{pmatrix}
    0 &0 &1 &0\\
    0 &0 &0 &0\\
    1 &0 &0 &0\\
    0 &0 &0 &0
\end{pmatrix},
\end{equation*}

\begin{equation*}
\lambda_5=
\begin{pmatrix}
    0 &0 &-i &0\\
    0 &0 &0 &0\\
    i &0 &0 &0\\
    0 &0 &0 &0
\end{pmatrix},
\lambda_6=
\begin{pmatrix}
    0 &0 &0 &1\\
    0 &0 &0 &0\\
    0 &0 &0 &0\\
    1 &0 &0 &0
\end{pmatrix},
\lambda_7=
\begin{pmatrix}
    0 &0 &0 &-i\\
    0 &0 &0 &0\\
    0 &0 &0 &0\\
    i &0 &0 &0
\end{pmatrix},
\end{equation*}

\begin{equation*}
\lambda_9=
\begin{pmatrix}
    0 &0 &0 &0\\
    0 &0 &1 &0\\
    0 &1 &0 &0\\
    0 &0 &0 &0
\end{pmatrix},
\lambda_{10}=
\begin{pmatrix}
    0 &0 &0 &0\\
    0 &0 &-i &0\\
    0 &i &0 &0\\
    0 &0 &0 &0
\end{pmatrix},
\lambda_{11}=
\begin{pmatrix}
    0 &0 &0 &0\\
    0 &0 &0 &1\\
    0 &0 &0 &0\\
    0 &1 &0 &0
\end{pmatrix},
\end{equation*}

\begin{equation*}
\lambda_{12}=
\begin{pmatrix}
    0 &0 &0 &0\\
    0 &0 &0 &-i\\
    0 &0 &0 &0\\
    0 &i &0 &0
\end{pmatrix},
\lambda_{13}=
\begin{pmatrix}
    0 &0 &0 &0\\
    0 &0 &0 &0\\
    0 &0 &0 &1\\
    0 &0 &1 &0
\end{pmatrix},
\lambda_{14}=
\begin{pmatrix}
    0 &0 &0 &0\\
    0 &0 &0 &0\\
    0 &0 &0 &-i\\
    0 &0 &i &0
\end{pmatrix},
\end{equation*}

\begin{equation*}
\lambda_3=
\begin{pmatrix}
    1 & & &\\
    &-1 & &\\
    & &0 &\\
    & & &0
\end{pmatrix},
\lambda_8=
\frac{1}{\sqrt3}
\begin{pmatrix}
    1 & & &\\
    &1 & &\\
    & &-2 &\\
    & & &0
\end{pmatrix},
\end{equation*}

\begin{equation}
\lambda_{15}=
\frac{1}{\sqrt6}
\begin{pmatrix}
    1 & & &\\
    &1 & &\\
    & &1 &\\
    & & &-3
\end{pmatrix}.
\end{equation}

They are the SU(4) generalization of the Gell-Mann matrices and satisfy 
\begin{equation} 
\lambda_a^\dagger=\lambda_a, \qquad \mathrm{Tr}\lambda_a=0 . 
\end{equation} 
With the normalization used here, one also has 
\begin{equation}
\mathrm{Tr}(\lambda_a\lambda_b)=2\delta_{ab}. 
\end{equation}
the SU(4) gauge potentials introduced in the main text can be written as

\begin{align}
A_x=&\frac{\pi}{4}(\lambda_1+\lambda_2+\lambda_4+\lambda_6-\lambda_7\nonumber\\
&+\lambda_9+\lambda_{10}+\lambda_{11}+\lambda_{13}+\lambda_{14}),\nonumber\\
A_y=&\frac{\pi}{4}(\lambda_3+\sqrt3 \lambda_8+\sqrt6 \lambda_{15}). 
\end{align}

It is also useful to identify the $\mathfrak{su}(2)$ algebra. We choose the basis to be the eigenbasis of $S_z$, the four-dimensional irreducible representation of the $\mathfrak{su}(2)$ generators is given by the spin-$3/2$ matrices
\begin{equation*}
S_x=
\begin{pmatrix}
0 & \dfrac{\sqrt{3}}{2} & 0 & 0 \\
\dfrac{\sqrt{3}}{2} & 0 & 1 & 0 \\
0 & 1 & 0 & \dfrac{\sqrt{3}}{2} \\
0 & 0 & \dfrac{\sqrt{3}}{2} & 0
\end{pmatrix},
\end{equation*}
\begin{equation*}
S_y=
\begin{pmatrix}
0 & -\dfrac{i\sqrt{3}}{2} & 0 & 0 \\
\dfrac{i\sqrt{3}}{2} & 0 & -i & 0 \\
0 & i & 0 & -\dfrac{i\sqrt{3}}{2} \\
0 & 0 & \dfrac{i\sqrt{3}}{2} & 0
\end{pmatrix},
\end{equation*}
\begin{equation}
S_z=
\begin{pmatrix}
\dfrac{3}{2} & 0 & 0 & 0 \\
0 & \dfrac{1}{2} & 0 & 0 \\
0 & 0 & -\dfrac{1}{2} & 0 \\
0 & 0 & 0 & -\dfrac{3}{2}
\end{pmatrix}.
\end{equation}
They satisfy the standard angular momentum algebra
\begin{equation}
    [S_\alpha,S_\beta]
    =
    i\epsilon_{\alpha\beta\gamma}S_\gamma,
    \qquad
    \alpha,\beta,\gamma=x,y,z .
\end{equation}
One can find the diagonal component of the non-Abelian gauge field used in the main text can be equivalently expressed as
\begin{equation}
    A_y = \frac{\pi}{2}S_z .
\end{equation}
This form makes clear that $A_y$ acts as a spin-dependent phase proportional to the spin-$3/2$ operator $S_z$, while $A_x$ contains off-diagonal components that mix different internal spin states.

\section{Explicit matrix forms of the Bloch Hamiltonians}
\label{Appendixb}

In this appendix, we present the explicit Bloch Hamiltonians for
the SU(4), SU(5), and SU(6) models
and give their general SU($N$) form. We use the ordered
internal-state basis
\begin{equation}
	\psi_{\mathbf{k}}
	=\bigl(c_{\mathbf{k},s},c_{\mathbf{k},s-1},\ldots,
	c_{\mathbf{k},-s}\bigr)^T,
	\qquad s=\frac{N-1}{2}.
\end{equation}
The Bloch Hamiltonian can be written as
\begin{equation}
	\label{eq:sun_bloch_links}
	\begin{aligned}
		\mathcal{H}_N(\mathbf{k})=-t\bigl[&
		e^{ik_x}e^{-iA_x^{(N)}}+e^{-ik_x}e^{iA_x^{(N)}}
		\\
		&+e^{ik_y}e^{-iA_y^{(N)}}+e^{-ik_y}e^{iA_y^{(N)}}
		\bigr].
	\end{aligned}
\end{equation}
For compactness, we introduce
\begin{equation}
	\label{eq:sun_abbreviations}
	\begin{gathered}
		\phi_N=\frac{(N-1)\pi}{N},
		\qquad q_N=e^{i(k_x-\phi_N)},
		\\
		c_\theta=2\cos(k_y-\theta).
	\end{gathered}
\end{equation}

For $N=4$, the gauge potentials specified in the main text yield
\begingroup
\setlength{\arraycolsep}{2pt}
\begin{equation}
	\label{eq:sun_h4}
	\mathcal{H}_4(\mathbf{k})=-t
	\begin{pmatrix}
		c_{3\pi/4} & q_4 & 0 & q_4^* \\
		q_4^* & c_{\pi/4} & q_4 & 0 \\
		0 & q_4^* & c_{-\pi/4} & q_4 \\
		q_4 & 0 & q_4^* & c_{-3\pi/4}
	\end{pmatrix}.
\end{equation}
Using the same gauge convention, the $N=5$ and $N=6$
Hamiltonians are
\begin{equation}
	\label{eq:sun_h5}
	\mathcal{H}_5(\mathbf{k})=-t
	\begin{pmatrix}
		c_{4\pi/5} & q_5 & 0 & 0 & q_5^* \\
		q_5^* & c_{2\pi/5} & q_5 & 0 & 0 \\
		0 & q_5^* & c_0 & q_5 & 0 \\
		0 & 0 & q_5^* & c_{-2\pi/5} & q_5 \\
		q_5 & 0 & 0 & q_5^* & c_{-4\pi/5}
	\end{pmatrix}
\end{equation}
and
\begin{equation}
	\label{eq:sun_h6}
	\mathcal{H}_6(\mathbf{k})=-t
	\begin{pmatrix}
		c_{5\pi/6} & q_6 & 0 & 0 & 0 & q_6^* \\
		q_6^* & c_{\pi/2} & q_6 & 0 & 0 & 0 \\
		0 & q_6^* & c_{\pi/6} & q_6 & 0 & 0 \\
		0 & 0 & q_6^* & c_{-\pi/6} & q_6 & 0 \\
		0 & 0 & 0 & q_6^* & c_{-\pi/2} & q_6 \\
		q_6 & 0 & 0 & 0 & q_6^* & c_{-5\pi/6}
	\end{pmatrix},
\end{equation}
\endgroup
respectively. Here the star denotes complex conjugation.
The corner entries close the cyclic coupling among the internal
states.

To obtain the general form, define
\begin{equation}
	\label{eq:sun_dn}
	D_N=\frac{\pi}{N}
	\operatorname{diag}(N-1,N-3,\ldots,1-N)
\end{equation}
and the discrete Fourier matrix
\begin{equation}
	\label{eq:sun_fourier}
	(F_N)_{j\ell}=\frac{1}{\sqrt{N}}
	e^{2\pi i(j-1)(\ell-1)/N},
\end{equation}
where $j,\ell=1,\ldots,N$. A uniform choice of gauge potentials is
\begin{equation}
	\label{eq:sun_gauge_potentials}
	A_y^{(N)}=D_N,
	\qquad
	A_x^{(N)}=F_ND_NF_N^\dagger.
\end{equation}
For $N=4$, this choice reproduces the gauge potentials in the main
text. Let $C_N$ be the cyclic shift matrix with elements
$(C_N)_{jk}=\delta^{(N)}_{k,j+1}$, where
$\delta^{(N)}_{a,b}=1$ if $a\equiv b\pmod N$ and is zero otherwise.
Writing $\omega_N=e^{2\pi i/N}$, we obtain
\begin{equation}
	\label{eq:sun_link_matrices}
	\begin{aligned}
		e^{-iA_x^{(N)}}&=e^{-i\phi_N}C_N,\\
		e^{-iA_y^{(N)}}&=e^{-i\phi_N}
		\operatorname{diag}(1,\omega_N,\ldots,\omega_N^{N-1}).
	\end{aligned}
\end{equation}
Substitution into Eq.~\eqref{eq:sun_bloch_links} gives
\begin{equation}
	\label{eq:sun_general_hamiltonian}
	\begin{aligned}
		\mathcal{H}_N(\mathbf{k})=-t\bigl[&
		q_NC_N+q_N^*C_N^\dagger
		\\
		&+2\cos(k_y {\zb\mathbb{I}_N}-D_N)\bigr],
	\end{aligned}
\end{equation}
where {\zb $\mathbb{I}_N$} is the $N\times N$ identity matrix. Equivalently,
the individual matrix elements are
\begin{equation}
	\label{eq:sun_general_elements}
	\begin{aligned}
		\bigl[\mathcal{H}_N(\mathbf{k})\bigr]_{jk}
		={}&-2t\cos\!\left[k_y-\frac{\pi}{N}(N+1-2j)\right]
		\delta_{jk}
		\\
		&-tq_N\delta^{(N)}_{k,j+1}
		-tq_N^*\delta^{(N)}_{k,j-1}.
	\end{aligned}
\end{equation}
Thus, for $N\geq3$, the Hamiltonian has a cyclic tridiagonal
structure: the diagonal cosine arguments advance by $2\pi/N$
between successive internal states, while the two cyclic hopping
amplitudes are $-tq_N$ and $-tq_N^*$.
Eq.~\eqref{eq:sun_general_elements} also applies to $N=2$,
where the two cyclic hopping contributions connect the same pair
of states and must be added.

Both gauge potentials in Eq.~\eqref{eq:sun_gauge_potentials}
are Hermitian and traceless, so their link matrices belong to
SU($N$). In particular, since
$\det C_N=(-1)^{N-1}$, the phase in
Eq.~\eqref{eq:sun_link_matrices} ensures
\begin{equation}
	\det\!\left(e^{-i\phi_N}C_N\right)
	=e^{-i(N-1)\pi}(-1)^{N-1}=1.
\end{equation}

\section{Pairing of spin-3/2 fermions\label{Appendixc}}
In this section, we summarize the local pairing channels of spin-$3/2$ fermions. For two spin-$3/2$ fermions, the total spin decomposition is
$\frac{3}{2} \otimes \frac{3}{2} = 0 \oplus 1 \oplus 2 \oplus 3$. For the onsite interaction considered in the main text, the orbital part of a Cooper pair is even under particle exchange. Therefore, the spin part must be antisymmetric. As a result, only the $J=0$ singlet channel and the $J=2$ quintet channel are allowed for onsite $s$-wave pairing {\zb (where 
$s$-wave indicates that the Cooper pair has zero orbital angular momentum)}, 
while the $J=1$ and $J=3$ channels are excluded by Fermi statistics. Possible onsite 
Cooper pairs can then be grouped into six distinct channels.

\noindent$J=0$ singlet {\zb Cooper} pairs:
\begin{align}
\ket{J=0, m=0}&=\frac12(\ket{\frac32, -\frac32}-\ket{-\frac32, \frac32}\nonumber\\&-\ket{\frac12, -\frac12}+\ket{-\frac12, \frac12}){\zb .}
\label{eq:J=0}
\end{align}

\noindent$J=2$ quintet {\zb Cooper} pairs:
\begin{align}
\ket{J=2, m=2}&=\frac{1}{\sqrt{2}}(\ket{\frac32, \frac12}-\ket{\frac12, \frac32}),\nonumber\\
\ket{J=2, m=1}&=\frac{1}{\sqrt{2}}(\ket{\frac32, -\frac12}-\ket{-\frac12, \frac32}),\nonumber\\
\ket{J=2, m=0}&=\frac12(\ket{\frac32, -\frac32}-\ket{-\frac32, \frac32}\nonumber\\&+\ket{\frac12, -\frac12}-\ket{-\frac12, \frac12}),\nonumber\\
\ket{J=2, m=-1}&=\frac{1}{\sqrt{2}}(\ket{-\frac32, \frac12}-\ket{\frac12, -\frac32}),\nonumber\\
\ket{J=2, m=-2}&=\frac{1}{\sqrt{2}}(\ket{-\frac32, -\frac12}-\ket{-\frac12, -\frac32}){\zb .}
\label{eq:J=2}
\end{align}
These six antisymmetric two-particle states provide the basis used to define the pairing operators in {\zb Sec.~\ref{IV}}. Explicitly, the pair creation operators are obtained by replacing the single-particle spin states in Eqs.~\ref{eq:J=0} and ~\ref{eq:J=2} with the corresponding fermion creation operators.

\bibliography{SUN}

@article{Neto2009graphene,
  title = {{The electronic properties of graphene}},
  author = {Castro Neto, A. H. and Guinea, F. and Peres, N. M. R. and Novoselov, K. S. and Geim, A. K.},
  journal = {Rev. Mod. Phys.},
  volume = {81},
  issue = {1},
  pages = {109--162},
  numpages = {0},
  year = {2009},
  month = {Jan},
  publisher = {American Physical Society},
  doi = {10.1103/RevModPhys.81.109},
  url = {https://link.aps.org/doi/10.1103/RevModPhys.81.109}
}

@article{Mo2025,
  title = {{Coexistence of chiral Majorana edge states and Bogoliubov Fermi surfaces in two-dimensional nonsymmorphic Dirac semimetal/superconductor heterostructures}},
  author = {Mo, Yijie and Wang, Xiao-Jiao and Zhuang, Zheng-Yang and Yan, Zhongbo},
  journal = {Phys. Rev. B},
  volume = {111},
  issue = {14},
  pages = {L140504},
  numpages = {7},
  year = {2025},
  month = {Apr},
  publisher = {American Physical Society},
  doi = {10.1103/PhysRevB.111.L140504},
  url = {https://link.aps.org/doi/10.1103/PhysRevB.111.L140504}
}

@article{Kopnin2013,
  title = {{High-temperature surface superconductivity in rhombohedral graphite}},
  author = {Kopnin, N. B. and Ij\"as, M. and Harju, A. and Heikkil\"a, T. T.},
  journal = {Phys. Rev. B},
  volume = {87},
  issue = {14},
  pages = {140503(R)},
  numpages = {4},
  year = {2013},
  month = {Apr},
  publisher = {American Physical Society},
  doi = {10.1103/PhysRevB.87.140503},
  url = {https://link.aps.org/doi/10.1103/PhysRevB.87.140503}
}

@article{Kopnin2011,
  title = {{High-temperature surface superconductivity in topological flat-band systems}},
  author = {Kopnin, N. B. and Heikkil\"a, T. T. and Volovik, G. E.},
  journal = {Phys. Rev. B},
  volume = {83},
  issue = {22},
  pages = {220503(R)},
  numpages = {4},
  year = {2011},
  month = {Jun},
  publisher = {American Physical Society},
  doi = {10.1103/PhysRevB.83.220503},
  url = {https://link.aps.org/doi/10.1103/PhysRevB.83.220503}
}

@article{Majid2023,
  title = {{Superconductivity in Luttinger semimetals near the SU(4) limit}},
  author = {Kheirkhah, Majid and Herbut, Igor F.},
  journal = {Phys. Rev. B},
  volume = {108},
  issue = {22},
  pages = {224514},
  numpages = {10},
  year = {2023},
  month = {Dec},
  publisher = {American Physical Society},
  doi = {10.1103/PhysRevB.108.224514},
  url = {https://link.aps.org/doi/10.1103/PhysRevB.108.224514}
}

@article{Liu2024SCS,
  title = {{Anomalous linear and quadratic nodeless surface Dirac cones in three-dimensional Dirac semimetals}},
  author = {Liu, Dongling and Wang, Xiao-Jiao and Mo, Yijie and Yan, Zhongbo},
  journal = {Phys. Rev. B},
  volume = {109},
  issue = {8},
  pages = {L081401},
  numpages = {7},
  year = {2024},
  month = {Feb},
  publisher = {American Physical Society},
  doi = {10.1103/PhysRevB.109.L081401},
  url = {https://link.aps.org/doi/10.1103/PhysRevB.109.L081401}
}

@article{Mo2024SCS,
  title = {{Boundary flat bands with topological spin textures protected by subchiral symmetry}},
  author = {Mo, Yijie and Wang, Xiao-Jiao and Yu, Rui and Yan, Zhongbo},
  journal = {Phys. Rev. B},
  volume = {109},
  issue = {24},
  pages = {245402},
  numpages = {12},
  year = {2024},
  month = {Jun},
  publisher = {American Physical Society},
  doi = {10.1103/PhysRevB.109.245402},
  url = {https://link.aps.org/doi/10.1103/PhysRevB.109.245402}
}

@article{Biao2024SCS,
  title = {{Experimental observation of boundary flat bands with topological spin textures}},
  author = {Biao, Yuanchuan and Yan, Zhongbo and Yu, Rui},
  journal = {Phys. Rev. B},
  volume = {110},
  issue = {24},
  pages = {L241110},
  numpages = {7},
  year = {2024},
  month = {Dec},
  publisher = {American Physical Society},
  doi = {10.1103/PhysRevB.110.L241110},
  url = {https://link.aps.org/doi/10.1103/PhysRevB.110.L241110}
}

@article{Hofstadter1976,
  title = {{Energy levels and wave functions of Bloch electrons in rational and irrational magnetic fields}},
  author = {Hofstadter, Douglas R.},
  journal = {Phys. Rev. B},
  volume = {14},
  issue = {6},
  pages = {2239--2249},
  numpages = {0},
  year = {1976},
  month = {Sep},
  publisher = {American Physical Society},
  doi = {10.1103/PhysRevB.14.2239},
  url = {https://link.aps.org/doi/10.1103/PhysRevB.14.2239}
}

@article{Radmanesh2018,
	title = {{Evidence for unconventional superconductivity in half-Heusler YPdBi and TbPdBi compounds revealed by London penetration depth measurements}},
	author = {Radmanesh, S. M. A. and Martin, C. and Zhu, Yanglin and Yin, X. and Xiao, H. and Mao, Z. Q. and Spinu, L.},
	journal = {Phys. Rev. B},
	volume = {98},
	issue = {24},
	pages = {241111(R)},
	numpages = {5},
	year = {2018},
	month = {Dec},
	publisher = {American Physical Society},
	doi = {10.1103/PhysRevB.98.241111},
	url = {https://link.aps.org/doi/10.1103/PhysRevB.98.241111}
}

@article{Jeong2021,
  title = {{${J}_{\mathrm{eff}}=\frac{3}{2}$ metallic phase and unconventional superconductivity in ${\mathrm{GaTa}}_{4}{\mathrm{Se}}_{8}$}},
  author = {Jeong, Min Yong and Chang, Seo Hyoung and Lee, Hyeong Jun and Sim, Jae-Hoon and Lee, Kyeong Jun and Janod, Etienne and Cario, Laurent and Said, Ayman and Bi, Wenli and Werner, Philipp and Go, Ara and Kim, Jungho and Han, Myung Joon},
  journal = {Phys. Rev. B},
  volume = {103},
  issue = {8},
  pages = {L081112},
  numpages = {6},
  year = {2021},
  month = {Feb},
  publisher = {American Physical Society},
  doi = {10.1103/PhysRevB.103.L081112},
  url = {https://link.aps.org/doi/10.1103/PhysRevB.103.L081112}
}

@article{Sigrist1991RMP,
	title = {{Phenomenological theory of unconventional superconductivity}},
	author = {Sigrist, Manfred and Ueda, Kazuo},
	journal = {Rev. Mod. Phys.},
	volume = {63},
	issue = {2},
	pages = {239--311},
	numpages = {0},
	year = {1991},
	month = {Apr},
	publisher = {American Physical Society},
	doi = {10.1103/RevModPhys.63.239},
	url = {https://link.aps.org/doi/10.1103/RevModPhys.63.239}
}

@article{Barnett2012,
	title = {{SU(3) Spin-Orbit Coupling in Systems of Ultracold Atoms}},
	author = {Barnett, Ryan and Boyd, G. R. and Galitski, Victor},
	journal = {Phys. Rev. Lett.},
	volume = {109},
	issue = {23},
	pages = {235308},
	numpages = {5},
	year = {2012},
	month = {Dec},
	publisher = {American Physical Society},
	doi = {10.1103/PhysRevLett.109.235308},
	url = {https://link.aps.org/doi/10.1103/PhysRevLett.109.235308}
}

@article{Wu2003,
	title = {{Exact SO(5) Symmetry in the Spin-$3/2$ Fermionic System}},
	author = {Wu, Congjun and Hu, Jiang-Ping and Zhang, Shou-Cheng},
	journal = {Phys. Rev. Lett.},
	volume = {91},
	issue = {18},
	pages = {186402},
	numpages = {4},
	year = {2003},
	month = {Oct},
	publisher = {American Physical Society},
	doi = {10.1103/PhysRevLett.91.186402},
	url = {https://link.aps.org/doi/10.1103/PhysRevLett.91.186402}
}

@article{Wu2006,
	author = {Wu, Congjun},
	title = {{Hidden symmetry and quantum phases and quantum phases in spin-$3/2$ cold atomic systems}},
	journal = {Modern Physics Letters B},
	volume = {20},
	number = {27},
	pages = {1707-1738},
	year = {2006},
	doi = {10.1142/S0217984906012213},
	URL = {https://doi.org/10.1142/S0217984906012213}
}

@article{Agterberg2017,
	title = {{Bogoliubov Fermi Surfaces in Superconductors with Broken Time-Reversal Symmetry}},
	author = {Agterberg, D. F. and Brydon, P. M. R. and Timm, C.},
	journal = {Phys. Rev. Lett.},
	volume = {118},
	issue = {12},
	pages = {127001},
	numpages = {6},
	year = {2017},
	month = {Mar},
	publisher = {American Physical Society},
	doi = {10.1103/PhysRevLett.118.127001},
	url = {https://link.aps.org/doi/10.1103/PhysRevLett.118.127001}
}

@article{Brydon2018,
	title = {{Bogoliubov Fermi surfaces: General theory, magnetic order, and topology}},
	author = {Brydon, P. M. R. and Agterberg, D. F. and Menke, Henri and Timm, C.},
	journal = {Phys. Rev. B},
	volume = {98},
	issue = {22},
	pages = {224509},
	numpages = {24},
	year = {2018},
	month = {Dec},
	publisher = {American Physical Society},
	doi = {10.1103/PhysRevB.98.224509},
	url = {https://link.aps.org/doi/10.1103/PhysRevB.98.224509}
}

@article{Dalibard2011,
	title = {{Colloquium: Artificial gauge potentials for neutral atoms}},
	author = {Dalibard, Jean and Gerbier, Fabrice and Juzeli\ifmmode \bar{u}\else \={u}\fi{}nas, Gediminas and \"Ohberg, Patrik},
	journal = {Rev. Mod. Phys.},
	volume = {83},
	issue = {4},
	pages = {1523--1543},
	numpages = {0},
	year = {2011},
	month = {Nov},
	publisher = {American Physical Society},
	doi = {10.1103/RevModPhys.83.1523},
	url = {https://link.aps.org/doi/10.1103/RevModPhys.83.1523}
}

@article{Juzeliunas2010,
	title = {{Generalized Rashba-Dresselhaus spin-orbit coupling for cold atoms}},
	author = {Juzeli\ifmmode \bar{u}\else \={u}\fi{}nas, Gediminas and Ruseckas, Julius and Dalibard, Jean},
	journal = {Phys. Rev. A},
	volume = {81},
	issue = {5},
	pages = {053403},
	numpages = {9},
	year = {2010},
	month = {May},
	publisher = {American Physical Society},
	doi = {10.1103/PhysRevA.81.053403},
	url = {https://link.aps.org/doi/10.1103/PhysRevA.81.053403}
}

@article{Venderbos2018,
	title = {{Pairing States of Spin-$\frac{3}{2}$ Fermions: Symmetry-Enforced Topological Gap Functions}},
	author = {Venderbos, J\"orn W. F. and Savary, Lucile and Ruhman, Jonathan and Lee, Patrick A. and Fu, Liang},
	journal = {Phys. Rev. X},
	volume = {8},
	issue = {1},
	pages = {011029},
	numpages = {31},
	year = {2018},
	month = {Feb},
	publisher = {American Physical Society},
	doi = {10.1103/PhysRevX.8.011029},
	url = {https://link.aps.org/doi/10.1103/PhysRevX.8.011029}
}

@article{Bornheimer2018,
	title = {{SU(3) topological insulators in the honeycomb lattice}},
	author = {Bornheimer, U. and Miniatura, C. and Gr\'emaud, B.},
	journal = {Phys. Rev. A},
	volume = {98},
	issue = {4},
	pages = {043614},
	numpages = {13},
	year = {2018},
	month = {Oct},
	publisher = {American Physical Society},
	doi = {10.1103/PhysRevA.98.043614},
	url = {https://link.aps.org/doi/10.1103/PhysRevA.98.043614}
}

@article{Brydon2016,
	title = {{Pairing of $j=3/2$ Fermions in Half-Heusler Superconductors}},
	author = {Brydon, P. M. R. and Wang, Limin and Weinert, M. and Agterberg, D. F.},
	journal = {Phys. Rev. Lett.},
	volume = {116},
	issue = {17},
	pages = {177001},
	numpages = {5},
	year = {2016},
	month = {Apr},
	publisher = {American Physical Society},
	doi = {10.1103/PhysRevLett.116.177001},
	url = {https://link.aps.org/doi/10.1103/PhysRevLett.116.177001}
}

@article{Dutta2021,
	title = {{Superconductivity in spin-$3/2$ systems: Symmetry classification, odd-frequency pairs, and Bogoliubov Fermi surfaces}},
	author = {Dutta, Paramita and Parhizgar, Fariborz and Black-Schaffer, Annica M.},
	journal = {Phys. Rev. Res.},
	volume = {3},
	issue = {3},
	pages = {033255},
	numpages = {13},
	year = {2021},
	month = {Sep},
	publisher = {American Physical Society},
	doi = {10.1103/PhysRevResearch.3.033255},
	url = {https://link.aps.org/doi/10.1103/PhysRevResearch.3.033255}
}

@article{
	Kim2018,
	author = {Hyunsoo Kim  and Kefeng Wang  and Yasuyuki Nakajima  and Rongwei Hu  and Steven Ziemak  and Paul Syers  and Limin Wang  and Halyna Hodovanets  and Jonathan D. Denlinger  and Philip M. R. Brydon  and Daniel F. Agterberg  and Makariy A. Tanatar  and Ruslan Prozorov  and Johnpierre Paglione },
	title = {{Beyond triplet: Unconventional superconductivity in a spin-3/2 topological semimetal}},
	journal = {Science Advances},
	volume = {4},
	number = {4},
	pages = {eaao4513},
	year = {2018},
	doi = {10.1126/sciadv.aao4513},
	URL = {https://www.science.org/doi/abs/10.1126/sciadv.aao4513}
}

@article{Ho1999,
	title = {{Pairing of Fermions with Arbitrary Spin}},
	author = {Ho, Tin-Lun and Yip, Sungkit},
	journal = {Phys. Rev. Lett.},
	volume = {82},
	issue = {2},
	pages = {247--250},
	numpages = {0},
	year = {1999},
	month = {Jan},
	publisher = {American Physical Society},
	doi = {10.1103/PhysRevLett.82.247},
	url = {https://link.aps.org/doi/10.1103/PhysRevLett.82.247}
}

@article{Hasan2010,
	title = {{Colloquium: Topological insulators}},
	author = {Hasan, M. Z. and Kane, C. L.},
	journal = {Rev. Mod. Phys.},
	volume = {82},
	issue = {4},
	pages = {3045--3067},
	numpages = {0},
	year = {2010},
	month = {Nov},
	publisher = {American Physical Society},
	doi = {10.1103/RevModPhys.82.3045},
	url = {https://link.aps.org/doi/10.1103/RevModPhys.82.3045}
}

@article{Qi2011,
	title = {{Topological insulators and superconductors}},
	author = {Qi, Xiao-Liang and Zhang, Shou-Cheng},
	journal = {Rev. Mod. Phys.},
	volume = {83},
	issue = {4},
	pages = {1057--1110},
	numpages = {0},
	year = {2011},
	month = {Oct},
	publisher = {American Physical Society},
	doi = {10.1103/RevModPhys.83.1057},
	url = {https://link.aps.org/doi/10.1103/RevModPhys.83.1057}
}

@article{Chiu2016,
	title = {{Classification of topological quantum matter with symmetries}},
	author = {Chiu, Ching-Kai and Teo, Jeffrey C. Y. and Schnyder, Andreas P. and Ryu, Shinsei},
	journal = {Rev. Mod. Phys.},
	volume = {88},
	issue = {3},
	pages = {035005},
	numpages = {63},
	year = {2016},
	month = {Aug},
	publisher = {American Physical Society},
	doi = {10.1103/RevModPhys.88.035005},
	url = {https://link.aps.org/doi/10.1103/RevModPhys.88.035005}
}

@article{Xiao2010,
	title = {{Berry phase effects on electronic properties}},
	author = {Xiao, Di and Chang, Ming-Che and Niu, Qian},
	journal = {Rev. Mod. Phys.},
	volume = {82},
	issue = {3},
	pages = {1959--2007},
	numpages = {0},
	year = {2010},
	month = {Jul},
	publisher = {American Physical Society},
	doi = {10.1103/RevModPhys.82.1959},
	url = {https://link.aps.org/doi/10.1103/RevModPhys.82.1959}
}

@article{thouless1982,
	title = {{Quantized Hall Conductance in a Two-Dimensional Periodic Potential}},
	author = {Thouless, D. J. and Kohmoto, M. and Nightingale, M. P. and den Nijs, M.},
	journal = {Phys. Rev. Lett.},
	volume = {49},
	issue = {6},
	pages = {405--408},
	numpages = {0},
	year = {1982},
	month = {Aug},
	publisher = {American Physical Society},
	doi = {10.1103/PhysRevLett.49.405},
	url = {https://link.aps.org/doi/10.1103/PhysRevLett.49.405}
}

@article{halperin1982,
	title = {{Quantized Hall conductance, current-carrying edge states, and the existence of extended states in a two-dimensional disordered potential}},
	author = {Halperin, B. I.},
	journal = {Phys. Rev. B},
	volume = {25},
	issue = {4},
	pages = {2185--2190},
	numpages = {0},
	year = {1982},
	month = {Feb},
	publisher = {American Physical Society},
	doi = {10.1103/PhysRevB.25.2185},
	url = {https://link.aps.org/doi/10.1103/PhysRevB.25.2185}
}

@article{haldane1988,
	title = {{Model for a Quantum Hall Effect without Landau Levels: Condensed-Matter Realization of the "Parity Anomaly"}},
	author = {Haldane, F. D. M.},
	journal = {Phys. Rev. Lett.},
	volume = {61},
	issue = {18},
	pages = {2015--2018},
	numpages = {0},
	year = {1988},
	month = {Oct},
	publisher = {American Physical Society},
	doi = {10.1103/PhysRevLett.61.2015},
	url = {https://link.aps.org/doi/10.1103/PhysRevLett.61.2015}
}

@article{hatsugai1993,
	title = {{Chern number and edge states in the integer quantum Hall effect}},
	author = {Hatsugai, Yasuhiro},
	journal = {Phys. Rev. Lett.},
	volume = {71},
	issue = {22},
	pages = {3697--3700},
	numpages = {0},
	year = {1993},
	month = {Nov},
	publisher = {American Physical Society},
	doi = {10.1103/PhysRevLett.71.3697},
	url = {https://link.aps.org/doi/10.1103/PhysRevLett.71.3697}
}

@article{kane2005,
	title = {{Quantum Spin Hall Effect in Graphene}},
	author = {Kane, C. L. and Mele, E. J.},
	journal = {Phys. Rev. Lett.},
	volume = {95},
	issue = {22},
	pages = {226801},
	numpages = {4},
	year = {2005},
	month = {Nov},
	publisher = {American Physical Society},
	doi = {10.1103/PhysRevLett.95.226801},
	url = {https://link.aps.org/doi/10.1103/PhysRevLett.95.226801}
}

@article{
	bernevig2006,
	author = {B. Andrei Bernevig  and Taylor L. Hughes  and Shou-Cheng Zhang },
	title = {{Quantum Spin Hall Effect and Topological Phase Transition in HgTe Quantum Wells}},
	journal = {Science},
	volume = {314},
	number = {5806},
	pages = {1757-1761},
	year = {2006},
	doi = {10.1126/science.1133734},
	URL = {https://www.science.org/doi/abs/10.1126/science.1133734}
}

@article{Nayak2008,
	title = {{Non-Abelian anyons and topological quantum computation}},
	author = {Nayak, Chetan and Simon, Steven H. and Stern, Ady and Freedman, Michael and Das Sarma, Sankar},
	journal = {Rev. Mod. Phys.},
	volume = {80},
	issue = {3},
	pages = {1083--1159},
	numpages = {0},
	year = {2008},
	month = {Sep},
	publisher = {American Physical Society},
	doi = {10.1103/RevModPhys.80.1083},
	url = {https://link.aps.org/doi/10.1103/RevModPhys.80.1083}
}

@article{bloch2008,
	title = {{Many-body physics with ultracold gases}},
	author = {Bloch, Immanuel and Dalibard, Jean and Zwerger, Wilhelm},
	journal = {Rev. Mod. Phys.},
	volume = {80},
	issue = {3},
	pages = {885--964},
	numpages = {0},
	year = {2008},
	month = {Jul},
	publisher = {American Physical Society},
	doi = {10.1103/RevModPhys.80.885},
	url = {https://link.aps.org/doi/10.1103/RevModPhys.80.885}
}

@article{lewenstein2007,
	author = {Maciej Lewenstein and Anna Sanpera and Veronica Ahufinger and Bogdan Damski and Aditi Sen(De) and Ujjwal Sen},
	title = {{Ultracold atomic gases in optical lattices: mimicking condensed matter physics and beyond}},
	journal = {Advances in Physics},
	volume = {56},
	number = {2},
	pages = {243--379},
	year = {2007},
	publisher = {Taylor \& Francis},
	doi = {10.1080/00018730701223200},
	URL = {https://doi.org/10.1080/00018730701223200}
}

@Article{bloch2012,
	author={Bloch, Immanuel
	and Dalibard, Jean
	and Nascimb{\`e}ne, Sylvain},
	title={{Quantum simulations with ultracold quantum gases}},
	journal={Nature Physics},
	year={2012},
	month={Apr},
	day={01},
	volume={8},
	number={4},
	pages={267-276},
	issn={1745-2481},
	doi={10.1038/nphys2259},
	url={https://doi.org/10.1038/nphys2259}
}

@article{lin2009,
	title = {{Bose-Einstein Condensate in a Uniform Light-Induced Vector Potential}},
	author = {Lin, Y.-J. and Compton, R. L. and Perry, A. R. and Phillips, W. D. and Porto, J. V. and Spielman, I. B.},
	journal = {Phys. Rev. Lett.},
	volume = {102},
	issue = {13},
	pages = {130401},
	numpages = {4},
	year = {2009},
	month = {Mar},
	publisher = {American Physical Society},
	doi = {10.1103/PhysRevLett.102.130401},
	url = {https://link.aps.org/doi/10.1103/PhysRevLett.102.130401}
}

@Article{lin2009Nature,
	author={Lin, Y.-J.
	and Compton, R. L.
	and Jim{\'e}nez-Garc{\'i}a, K.
	and Porto, J. V.
	and Spielman, I. B.},
	title={{Synthetic magnetic fields for ultracold neutral atoms}},
	journal={Nature},
	year={2009},
	month={Dec},
	day={01},
	volume={462},
	number={7273},
	pages={628-632},
	issn={1476-4687},
	doi={10.1038/nature08609},
	url={https://doi.org/10.1038/nature08609}
}

@Article{lin2011,
	author={Lin, Y.-J.
	and Jim{\'e}nez-Garc{\'i}a, K.
	and Spielman, I. B.},
	title={{Spin--orbit-coupled Bose--Einstein condensates}},
	journal={Nature},
	year={2011},
	month={Mar},
	day={01},
	volume={471},
	number={7336},
	pages={83-86},
	issn={1476-4687},
	doi={10.1038/nature09887},
	url={https://doi.org/10.1038/nature09887}
}

@article{goldman2014,
	doi = {10.1088/0034-4885/77/12/126401},
	url = {https://doi.org/10.1088/0034-4885/77/12/126401},
	year = {2014},
	month = {nov},
	publisher = {IOP Publishing},
	volume = {77},
	number = {12},
	pages = {126401},
	author = {Goldman, N and Juzeliūnas, G and Öhberg, P and Spielman, I B},
	title = {{Light-induced gauge fields for ultracold atoms}},
	journal = {Reports on Progress in Physics}
}

@article{cooper2019,
	title = {{Topological bands for ultracold atoms}},
	author = {Cooper, N. R. and Dalibard, J. and Spielman, I. B.},
	journal = {Rev. Mod. Phys.},
	volume = {91},
	issue = {1},
	pages = {015005},
	numpages = {55},
	year = {2019},
	month = {Mar},
	publisher = {American Physical Society},
	doi = {10.1103/RevModPhys.91.015005},
	url = {https://link.aps.org/doi/10.1103/RevModPhys.91.015005}
}

@article{
	gross2017,
	author = {Christian Gross  and Immanuel Bloch },
	title = {{Quantum simulations with ultracold atoms in optical lattices}},
	journal = {Science},
	volume = {357},
	number = {6355},
	pages = {995-1001},
	year = {2017},
	doi = {10.1126/science.aal3837},
	URL = {https://www.science.org/doi/abs/10.1126/science.aal3837}
}

@article{zhai2015,
	doi = {10.1088/0034-4885/78/2/026001},
	url = {https://doi.org/10.1088/0034-4885/78/2/026001},
	year = {2015},
	month = {feb},
	publisher = {IOP Publishing},
	volume = {78},
	number = {2},
	pages = {026001},
	author = {Zhai, Hui},
	title = {{Degenerate quantum gases with spin–orbit coupling: a review}},
	journal = {Reports on Progress in Physics}
}

@Article{galitski2013,
	author={Galitski, Victor
	and Spielman, Ian B.},
	title={{Spin--orbit coupling in quantum gases}},
	journal={Nature},
	year={2013},
	month={Feb},
	day={01},
	volume={494},
	number={7435},
	pages={49-54},
	issn={1476-4687},
	doi={10.1038/nature11841},
	url={https://doi.org/10.1038/nature11841}
}

@article{wang2012,
	title = {{Spin-Orbit Coupled Degenerate Fermi Gases}},
	author = {Wang, Pengjun and Yu, Zeng-Qiang and Fu, Zhengkun and Miao, Jiao and Huang, Lianghui and Chai, Shijie and Zhai, Hui and Zhang, Jing},
	journal = {Phys. Rev. Lett.},
	volume = {109},
	issue = {9},
	pages = {095301},
	numpages = {5},
	year = {2012},
	month = {Aug},
	publisher = {American Physical Society},
	doi = {10.1103/PhysRevLett.109.095301},
	url = {https://link.aps.org/doi/10.1103/PhysRevLett.109.095301}
}

@article{cheuk2012,
	title = {{Spin-Injection Spectroscopy of a Spin-Orbit Coupled Fermi Gas}},
	author = {Cheuk, Lawrence W. and Sommer, Ariel T. and Hadzibabic, Zoran and Yefsah, Tarik and Bakr, Waseem S. and Zwierlein, Martin W.},
	journal = {Phys. Rev. Lett.},
	volume = {109},
	issue = {9},
	pages = {095302},
	numpages = {5},
	year = {2012},
	month = {Aug},
	publisher = {American Physical Society},
	doi = {10.1103/PhysRevLett.109.095302},
	url = {https://link.aps.org/doi/10.1103/PhysRevLett.109.095302}
}

@article{aidelsburger2011,
	title = {{Experimental Realization of Strong Effective Magnetic Fields in an Optical Lattice}},
	author = {Aidelsburger, M. and Atala, M. and Nascimb\`ene, S. and Trotzky, S. and Chen, Y.-A. and Bloch, I.},
	journal = {Phys. Rev. Lett.},
	volume = {107},
	issue = {25},
	pages = {255301},
	numpages = {5},
	year = {2011},
	month = {Dec},
	publisher = {American Physical Society},
	doi = {10.1103/PhysRevLett.107.255301},
	url = {https://link.aps.org/doi/10.1103/PhysRevLett.107.255301}
}

@article{aidelsburger2013,
  title = {{Realization of the Hofstadter Hamiltonian with Ultracold Atoms in Optical Lattices}},
  author = {Aidelsburger, M. and Atala, M. and Lohse, M. and Barreiro, J. T. and Paredes, B. and Bloch, I.},
  journal = {Phys. Rev. Lett.},
  volume = {111},
  issue = {18},
  pages = {185301},
  numpages = {5},
  year = {2013},
  month = {Oct},
  publisher = {American Physical Society},
  doi = {10.1103/PhysRevLett.111.185301},
  url = {https://link.aps.org/doi/10.1103/PhysRevLett.111.185301}
}

@Article{gorshkov2010,
	author={Gorshkov, A. V.
	and Hermele, M.
	and Gurarie, V.
	and Xu, C.
	and Julienne, P. S.
	and Ye, J.
	and Zoller, P.
	and Demler, E.
	and Lukin, M. D.
	and Rey, A. M.},
	title={{Two-orbital SU($N$) magnetism with ultracold alkaline-earth atoms}},
	journal={Nature Physics},
	year={2010},
	month={Apr},
	day={01},
	volume={6},
	number={4},
	pages={289-295},
	issn={1745-2481},
	doi={10.1038/nphys1535},
	url={https://doi.org/10.1038/nphys1535}
}

@article{cazalilla2014,
	doi = {10.1088/0034-4885/77/12/124401},
	url = {https://doi.org/10.1088/0034-4885/77/12/124401},
	year = {2014},
	month = {nov},
	publisher = {IOP Publishing},
	volume = {77},
	number = {12},
	pages = {124401},
	author = {Cazalilla, Miguel A and Rey, Ana Maria},
	title = {{Ultracold Fermi gases with emergent SU($N$) symmetry}},
	journal = {Reports on Progress in Physics},
}

@Article{taie2012,
	author={Taie, Shintaro
	and Yamazaki, Rekishu
	and Sugawa, Seiji
	and Takahashi, Yoshiro},
	title={{An SU(6) Mott insulator of an atomic Fermi gas realized by large-spin Pomeranchuk cooling}},
	journal={Nature Physics},
	year={2012},
	month={Nov},
	day={01},
	volume={8},
	number={11},
	pages={825-830},
	issn={1745-2481},
	doi={10.1038/nphys2430},
	url={https://doi.org/10.1038/nphys2430}
}

@Article{scazza2014,
	author={Scazza, F.
	and Hofrichter, C.
	and H{\"o}fer, M.
	and De Groot, P. C.
	and Bloch, I.
	and F{\"o}lling, S.},
	title={{Observation of two-orbital spin-exchange interactions with ultracold SU(N)-symmetric fermions}},
	journal={Nature Physics},
	year={2014},
	month={Oct},
	day={01},
	volume={10},
	number={10},
	pages={779-784},
	issn={1745-2481},
	doi={10.1038/nphys3061},
	url={https://doi.org/10.1038/nphys3061}
}

@article{
	zhang2014,
	author = {X. Zhang  and M. Bishof  and S. L. Bromley  and C. V. Kraus  and M. S. Safronova  and P. Zoller  and A. M. Rey  and J. Ye },
	title = {{Spectroscopic observation of SU($N$)-symmetric interactions in Sr orbital magnetism}},
	journal = {Science},
	volume = {345},
	number = {6203},
	pages = {1467-1473},
	year = {2014},
	doi = {10.1126/science.1254978},
	URL = {https://www.science.org/doi/abs/10.1126/science.1254978}
}

@Article{pagano2014,
	author={Pagano, Guido
	and Mancini, Marco
	and Cappellini, Giacomo
	and Lombardi, Pietro
	and Sch{\"a}fer, Florian
	and Hu, Hui
	and Liu, Xia-Ji
	and Catani, Jacopo
	and Sias, Carlo
	and Inguscio, Massimo
	and Fallani, Leonardo},
	title={{A one-dimensional liquid of fermions with tunable spin}},
	journal={Nature Physics},
	year={2014},
	month={Mar},
	day={01},
	volume={10},
	number={3},
	pages={198-201},
	issn={1745-2481},
	doi={10.1038/nphys2878},
	url={https://doi.org/10.1038/nphys2878}
}

@article{yip1999,
	title = {{Zero sound modes of dilute Fermi gases with arbitrary spin}},
	author = {Yip, S.-K. and Ho, Tin-Lun},
	journal = {Phys. Rev. A},
	volume = {59},
	issue = {6},
	pages = {4653--4656},
	numpages = {0},
	year = {1999},
	month = {Jun},
	publisher = {American Physical Society},
	doi = {10.1103/PhysRevA.59.4653},
	url = {https://link.aps.org/doi/10.1103/PhysRevA.59.4653}
}

@article{Yang2016,
	title = {{Topological Septet Pairing with Spin-$\frac{3}{2}$ Fermions: High-Partial-Wave Channel Counterpart of the $^{3}\mathrm{He}\text{\ensuremath{-}}B$ Phase}},
	author = {Yang, Wang and Li, Yi and Wu, Congjun},
	journal = {Phys. Rev. Lett.},
	volume = {117},
	issue = {7},
	pages = {075301},
	numpages = {6},
	year = {2016},
	month = {Aug},
	publisher = {American Physical Society},
	doi = {10.1103/PhysRevLett.117.075301},
	url = {https://link.aps.org/doi/10.1103/PhysRevLett.117.075301}
}

@article{Roy2019,
	title = {{Topological superconductivity of spin-$3/2$ carriers in a three-dimensional doped Luttinger semimetal}},
	author = {Roy, Bitan and Ghorashi, Sayed Ali Akbar and Foster, Matthew S. and Nevidomskyy, Andriy H.},
	journal = {Phys. Rev. B},
	volume = {99},
	issue = {5},
	pages = {054505},
	numpages = {44},
	year = {2019},
	month = {Feb},
	publisher = {American Physical Society},
	doi = {10.1103/PhysRevB.99.054505},
	url = {https://link.aps.org/doi/10.1103/PhysRevB.99.054505}
}

@article{boettcher2016,
	title = {{Superconducting quantum criticality in three-dimensional Luttinger semimetals}},
	author = {Boettcher, Igor and Herbut, Igor F.},
	journal = {Phys. Rev. B},
	volume = {93},
	issue = {20},
	pages = {205138},
	numpages = {16},
	year = {2016},
	month = {May},
	publisher = {American Physical Society},
	doi = {10.1103/PhysRevB.93.205138},
	url = {https://link.aps.org/doi/10.1103/PhysRevB.93.205138}
}

@article{boettcher2018,
	title = {{Unconventional Superconductivity in Luttinger Semimetals: Theory of Complex Tensor Order and the Emergence of the Uniaxial Nematic State}},
	author = {Boettcher, Igor and Herbut, Igor F.},
	journal = {Phys. Rev. Lett.},
	volume = {120},
	issue = {5},
	pages = {057002},
	numpages = {6},
	year = {2018},
	month = {Feb},
	publisher = {American Physical Society},
	doi = {10.1103/PhysRevLett.120.057002},
	url = {https://link.aps.org/doi/10.1103/PhysRevLett.120.057002}
}

@article{butch2011,
	title = {{Superconductivity in the topological semimetal YPtBi}},
	author = {Butch, N. P. and Syers, P. and Kirshenbaum, K. and Hope, A. P. and Paglione, J.},
	journal = {Phys. Rev. B},
	volume = {84},
	issue = {22},
	pages = {220504(R)},
	numpages = {5},
	year = {2011},
	month = {Dec},
	publisher = {American Physical Society},
	doi = {10.1103/PhysRevB.84.220504},
	url = {https://link.aps.org/doi/10.1103/PhysRevB.84.220504}
}

@article{bay2012,
	title = {{Superconductivity in noncentrosymmetric YPtBi under pressure}},
	author = {Bay, T. V. and Naka, T. and Huang, Y. K. and de Visser, A.},
	journal = {Phys. Rev. B},
	volume = {86},
	issue = {6},
	pages = {064515},
	numpages = {5},
	year = {2012},
	month = {Aug},
	publisher = {American Physical Society},
	doi = {10.1103/PhysRevB.86.064515},
	url = {https://link.aps.org/doi/10.1103/PhysRevB.86.064515}
}

@article{tafti2013,
	title = {{Superconductivity in the noncentrosymmetric half-Heusler compound LuPtBi: A candidate for topological superconductivity}},
	author = {Tafti, F. F. and Fujii, Takenori and Juneau-Fecteau, A. and Ren\'e de Cotret, S. and Doiron-Leyraud, N. and Asamitsu, Atsushi and Taillefer, Louis},
	journal = {Phys. Rev. B},
	volume = {87},
	issue = {18},
	pages = {184504},
	numpages = {5},
	year = {2013},
	month = {May},
	publisher = {American Physical Society},
	doi = {10.1103/PhysRevB.87.184504},
	url = {https://link.aps.org/doi/10.1103/PhysRevB.87.184504}
}

@article{
	nakajima2015,
	author = {Yasuyuki Nakajima  and Rongwei Hu  and Kevin Kirshenbaum  and Alex Hughes  and Paul Syers  and Xiangfeng Wang  and Kefeng Wang  and Renxiong Wang  and Shanta R. Saha  and Daniel Pratt  and Jeffrey W. Lynn  and Johnpierre Paglione },
	title = {{Topological {$R$}PdBi half-Heusler semimetals: A new family of noncentrosymmetric magnetic superconductors}},
	journal = {Science Advances},
	volume = {1},
	number = {5},
	pages = {e1500242},
	year = {2015},
	doi = {10.1126/sciadv.1500242},
	URL = {https://www.science.org/doi/abs/10.1126/sciadv.1500242}
}

@article{meinert2016,
	title = {{Unconventional Superconductivity in YPtBi and Related Topological Semimetals}},
	author = {Meinert, Markus},
	journal = {Phys. Rev. Lett.},
	volume = {116},
	issue = {13},
	pages = {137001},
	numpages = {5},
	year = {2016},
	month = {Apr},
	publisher = {American Physical Society},
	doi = {10.1103/PhysRevLett.116.137001},
	url = {https://link.aps.org/doi/10.1103/PhysRevLett.116.137001}
}

@article{pan2013,
	doi = {10.1209/0295-5075/104/27001},
	url = {https://doi.org/10.1209/0295-5075/104/27001},
	year = {2013},
	month = {oct},
	publisher = {EDP Sciences, IOP Publishing and Società Italiana di Fisica},
	volume = {104},
	number = {2},
	pages = {27001},
	author = {Pan, Y. and Nikitin, A. M. and Bay, T. V. and Huang, Y. K. and Paulsen, C. and Yan, B. H. and de Visser, A.},
	title = {{Superconductivity and magnetic order in the noncentrosymmetric half-Heusler compound ErPdBi}},
	journal = {Europhysics Letters}
}

@article{menke2019,
	title = {{Bogoliubov Fermi surfaces stabilized by spin-orbit coupling}},
	author = {Menke, Henri and Timm, C. and Brydon, P. M. R.},
	journal = {Phys. Rev. B},
	volume = {100},
	issue = {22},
	pages = {224505},
	numpages = {12},
	year = {2019},
	month = {Dec},
	publisher = {American Physical Society},
	doi = {10.1103/PhysRevB.100.224505},
	url = {https://link.aps.org/doi/10.1103/PhysRevB.100.224505}
}

@article{link2020,
	title = {{$d$-wave superconductivity and Bogoliubov-Fermi surfaces in Rarita-Schwinger-Weyl semimetals}},
	author = {Link, Julia M. and Boettcher, Igor and Herbut, Igor F.},
	journal = {Phys. Rev. B},
	volume = {101},
	issue = {18},
	pages = {184503},
	numpages = {19},
	year = {2020},
	month = {May},
	publisher = {American Physical Society},
	doi = {10.1103/PhysRevB.101.184503},
	url = {https://link.aps.org/doi/10.1103/PhysRevB.101.184503}
}

@article{yuan2018,
	title = {{Zeeman-induced gapless superconductivity with a partial Fermi surface}},
	author = {Yuan, Noah F. Q. and Fu, Liang},
	journal = {Phys. Rev. B},
	volume = {97},
	issue = {11},
	pages = {115139},
	numpages = {5},
	year = {2018},
	month = {Mar},
	publisher = {American Physical Society},
	doi = {10.1103/PhysRevB.97.115139},
	url = {https://link.aps.org/doi/10.1103/PhysRevB.97.115139}
}

@article{lapp2020,
	title = {{Experimental consequences of Bogoliubov Fermi surfaces}},
	author = {Lapp, Clara J. and B\"orner, Georg and Timm, Carsten},
	journal = {Phys. Rev. B},
	volume = {101},
	issue = {2},
	pages = {024505},
	numpages = {16},
	year = {2020},
	month = {Jan},
	publisher = {American Physical Society},
	doi = {10.1103/PhysRevB.101.024505},
	url = {https://link.aps.org/doi/10.1103/PhysRevB.101.024505}
}

@Article{setty2020,
	author={Setty, Chandan
	and Bhattacharyya, Shinibali
	and Cao, Yifu
	and Kreisel, Andreas
	and Hirschfeld, P. J.},
	title={{Topological ultranodal pair states in iron-based superconductors}},
	journal={Nature Communications},
	year={2020},
	month={Jan},
	day={27},
	volume={11},
	number={1},
	pages={523},
	issn={2041-1723},
	doi={10.1038/s41467-020-14357-2},
	url={https://doi.org/10.1038/s41467-020-14357-2}
}

@article{tang2011,
	title = {{High-Temperature Fractional Quantum Hall States}},
	author = {Tang, Evelyn and Mei, Jia-Wei and Wen, Xiao-Gang},
	journal = {Phys. Rev. Lett.},
	volume = {106},
	issue = {23},
	pages = {236802},
	numpages = {4},
	year = {2011},
	month = {Jun},
	publisher = {American Physical Society},
	doi = {10.1103/PhysRevLett.106.236802},
	url = {https://link.aps.org/doi/10.1103/PhysRevLett.106.236802}
}

@article{sun2011,
	title ={{Nearly Flatbands with Nontrivial Topology}},
	author = {Sun, Kai and Gu, Zhengcheng and Katsura, Hosho and Das Sarma, S.},
	journal = {Phys. Rev. Lett.},
	volume = {106},
	issue = {23},
	pages = {236803},
	numpages = {4},
	year = {2011},
	month = {Jun},
	publisher = {American Physical Society},
	doi = {10.1103/PhysRevLett.106.236803},
	url = {https://link.aps.org/doi/10.1103/PhysRevLett.106.236803}
}

@article{neupert2011,
	title = {{Fractional Quantum Hall States at Zero Magnetic Field}},
	author = {Neupert, Titus and Santos, Luiz and Chamon, Claudio and Mudry, Christopher},
	journal = {Phys. Rev. Lett.},
	volume = {106},
	issue = {23},
	pages = {236804},
	numpages = {4},
	year = {2011},
	month = {Jun},
	publisher = {American Physical Society},
	doi = {10.1103/PhysRevLett.106.236804},
	url = {https://link.aps.org/doi/10.1103/PhysRevLett.106.236804}
}

@Article{sheng2011,
	author={Sheng, D. N.
	and Gu, Zheng-Cheng
	and Sun, Kai
	and Sheng, L.},
	title={{Fractional quantum Hall effect in the absence of Landau levels}},
	journal={Nature Communications},
	year={2011},
	month={Jul},
	day={12},
	volume={2},
	number={1},
	pages={389},
	issn={2041-1723},
	doi={10.1038/ncomms1380},
	url={https://doi.org/10.1038/ncomms1380}
}

@article{regnault2011,
	title = {{Fractional Chern Insulator}},
	author = {Regnault, N. and Bernevig, B. Andrei},
	journal = {Phys. Rev. X},
	volume = {1},
	issue = {2},
	pages = {021014},
	numpages = {14},
	year = {2011},
	month = {Dec},
	publisher = {American Physical Society},
	doi = {10.1103/PhysRevX.1.021014},
	url = {https://link.aps.org/doi/10.1103/PhysRevX.1.021014}
}

@article{bergholtz2013,
	author = {Bergholtz, Emil J. and Liu, Zhao},
	title = {{Topological flat band models and fractional Chern insulators}},
	journal = {International Journal of Modern Physics B},
	volume = {27},
	number = {24},
	pages = {1330017},
	year = {2013},
	doi = {10.1142/S021797921330017X},
	URL = {https://doi.org/10.1142/S021797921330017X}
}

@article{parameswaran2013,
	author = {Siddharth A. Parameswaran and Rahul Roy and Shivaji L. Sondhi},
	title = {{Fractional quantum Hall physics in topological flat bands}},
	journal = {Comptes Rendus. Physique},
	pages = {816--839},
	year = {2013},
	publisher = {Elsevier},
	volume = {14},
	number = {9-10},
	doi = {10.1016/j.crhy.2013.04.003}
}

@article{roy2014,
	title = {{Band geometry of fractional topological insulators}},
	author = {Roy, Rahul},
	journal = {Phys. Rev. B},
	volume = {90},
	issue = {16},
	pages = {165139},
	numpages = {7},
	year = {2014},
	month = {Oct},
	publisher = {American Physical Society},
	doi = {10.1103/PhysRevB.90.165139},
	url = {https://link.aps.org/doi/10.1103/PhysRevB.90.165139}
}

@Article{jackson2015,
	author={Jackson, T. S.
	and M{\"o}ller, Gunnar
	and Roy, Rahul},
	title={{Geometric stability of topological lattice phases}},
	journal={Nature Communications},
	year={2015},
	month={Nov},
	day={04},
	volume={6},
	number={1},
	pages={8629},
	issn={2041-1723},
	doi={10.1038/ncomms9629},
	url={https://doi.org/10.1038/ncomms9629}
}

@article{yang2012,
	title = {{Topological flat band models with arbitrary Chern numbers}},
	author = {Yang, Shuo and Gu, Zheng-Cheng and Sun, Kai and Das Sarma, S.},
	journal = {Phys. Rev. B},
	volume = {86},
	issue = {24},
	pages = {241112(R)},
	numpages = {5},
	year = {2012},
	month = {Dec},
	publisher = {American Physical Society},
	doi = {10.1103/PhysRevB.86.241112},
	url = {https://link.aps.org/doi/10.1103/PhysRevB.86.241112}
}

@article{liu2012,
	title = {{Fractional Chern Insulators in Topological Flat Bands with Higher Chern Number}},
	author = {Liu, Zhao and Bergholtz, Emil J. and Fan, Heng and L\"auchli, Andreas M.},
	journal = {Phys. Rev. Lett.},
	volume = {109},
	issue = {18},
	pages = {186805},
	numpages = {5},
	year = {2012},
	month = {Nov},
	publisher = {American Physical Society},
	doi = {10.1103/PhysRevLett.109.186805},
	url = {https://link.aps.org/doi/10.1103/PhysRevLett.109.186805}
}

@Article{Cai2023,
	author={Cai, Jiaqi
	and Anderson, Eric
	and Wang, Chong
	and Zhang, Xiaowei
	and Liu, Xiaoyu
	and Holtzmann, William
	and Zhang, Yinong
	and Fan, Fengren
	and Taniguchi, Takashi
	and Watanabe, Kenji
	and Ran, Ying
	and Cao, Ting
	and Fu, Liang
	and Xiao, Di
	and Yao, Wang
	and Xu, Xiaodong},
	title={{Signatures of fractional quantum anomalous Hall states in twisted MoTe2}},
	journal={Nature},
	year={2023},
	month={Oct},
	day={01},
	volume={622},
	number={7981},
	pages={63-68},
	issn={1476-4687},
	doi={10.1038/s41586-023-06289-w},
	url={https://doi.org/10.1038/s41586-023-06289-w}
}

@Article{Zeng2023,
	author={Zeng, Yihang
	and Xia, Zhengchao
	and Kang, Kaifei
	and Zhu, Jiacheng
	and Kn{\"u}ppel, Patrick
	and Vaswani, Chirag
	and Watanabe, Kenji
	and Taniguchi, Takashi
	and Mak, Kin Fai
	and Shan, Jie},
	title={{Thermodynamic evidence of fractional Chern insulator in moir{\'e} MoTe2}},
	journal={Nature},
	year={2023},
	month={Oct},
	day={01},
	volume={622},
	number={7981},
	pages={69-73},
	issn={1476-4687},
	doi={10.1038/s41586-023-06452-3},
	url={https://doi.org/10.1038/s41586-023-06452-3}
}

@Article{Park2023,
	author={Park, Heonjoon
	and Cai, Jiaqi
	and Anderson, Eric
	and Zhang, Yinong
	and Zhu, Jiayi
	and Liu, Xiaoyu
	and Wang, Chong
	and Holtzmann, William
	and Hu, Chaowei
	and Liu, Zhaoyu
	and Taniguchi, Takashi
	and Watanabe, Kenji
	and Chu, Jiun-Haw
	and Cao, Ting
	and Fu, Liang
	and Yao, Wang
	and Chang, Cui-Zu
	and Cobden, David
	and Xiao, Di
	and Xu, Xiaodong},
	title={{Observation of fractionally quantized anomalous Hall effect}},
	journal={Nature},
	year={2023},
	month={Oct},
	day={01},
	volume={622},
	number={7981},
	pages={74-79},
	issn={1476-4687},
	doi={10.1038/s41586-023-06536-0},
	url={https://doi.org/10.1038/s41586-023-06536-0}
}

@article{Ledwith2023,
	title = {{Vortexability: A unifying criterion for ideal fractional Chern insulators}},
	author = {Ledwith, Patrick J. and Vishwanath, Ashvin and Parker, Daniel E.},
	journal = {Phys. Rev. B},
	volume = {108},
	issue = {20},
	pages = {205144},
	numpages = {20},
	year = {2023},
	month = {Nov},
	publisher = {American Physical Society},
	doi = {10.1103/PhysRevB.108.205144},
	url = {https://link.aps.org/doi/10.1103/PhysRevB.108.205144}
}

@Article{Taie2022,
	author={Taie, Shintaro
	and Ibarra-Garc{\'i}a-Padilla, Eduardo
	and Nishizawa, Naoki
	and Takasu, Yosuke
	and Kuno, Yoshihito
	and Wei, Hao-Tian
	and Scalettar, Richard T.
	and Hazzard, Kaden R. A.
	and Takahashi, Yoshiro},
	title={{Observation of antiferromagnetic correlations in an ultracold SU(N) Hubbard model}},
	journal={Nature Physics},
	year={2022},
	month={Nov},
	day={01},
	volume={18},
	number={11},
	pages={1356-1361},
	issn={1745-2481},
	doi={10.1038/s41567-022-01725-6},
	url={https://doi.org/10.1038/s41567-022-01725-6}
}

@article{Pasqualetti2024,
	title = {{Equation of State and Thermometry of the 2D $\mathrm{SU}(N)$ Fermi-Hubbard Model}},
	author = {Pasqualetti, G. and Bettermann, O. and Darkwah Oppong, N. and Ibarra-Garc\'{\i}a-Padilla, E. and Dasgupta, S. and Scalettar, R. T. and Hazzard, K. R. A. and Bloch, I. and F\"olling, S.},
	journal = {Phys. Rev. Lett.},
	volume = {132},
	issue = {8},
	pages = {083401},
	numpages = {8},
	year = {2024},
	month = {Feb},
	publisher = {American Physical Society},
	doi = {10.1103/PhysRevLett.132.083401},
	url = {https://link.aps.org/doi/10.1103/PhysRevLett.132.083401}
}

@article{Ruseckas2005,
	title = {{Non-Abelian Gauge Potentials for Ultracold Atoms with Degenerate Dark States}},
	author = {Ruseckas, J. and Juzeli\ifmmode \bar{u}\else \={u}\fi{}nas, G. and \"Ohberg, P. and Fleischhauer, M.},
	journal = {Phys. Rev. Lett.},
	volume = {95},
	issue = {1},
	pages = {010404},
	numpages = {4},
	year = {2005},
	month = {Jun},
	publisher = {American Physical Society},
	doi = {10.1103/PhysRevLett.95.010404},
	url = {https://link.aps.org/doi/10.1103/PhysRevLett.95.010404}
}

@article{Burrello2011,
	title = {{Ultracold atoms in U(2) non-Abelian gauge potentials preserving the Landau levels}},
	author = {Burrello, Michele and Trombettoni, Andrea},
	journal = {Phys. Rev. A},
	volume = {84},
	issue = {4},
	pages = {043625},
	numpages = {18},
	year = {2011},
	month = {Oct},
	publisher = {American Physical Society},
	doi = {10.1103/PhysRevA.84.043625},
	url = {https://link.aps.org/doi/10.1103/PhysRevA.84.043625}
}

@article{Hasan2022,
	title = {{Wave Packet Dynamics in Synthetic Non-Abelian Gauge Fields}},
	author = {Hasan, Mehedi and Madasu, Chetan Sriram and Rathod, Ketan D. and Kwong, Chang Chi and Miniatura, Christian and Chevy, Fr\'ed\'eric and Wilkowski, David},
	journal = {Phys. Rev. Lett.},
	volume = {129},
	issue = {13},
	pages = {130402},
	numpages = {6},
	year = {2022},
	month = {Sep},
	publisher = {American Physical Society},
	doi = {10.1103/PhysRevLett.129.130402},
	url = {https://link.aps.org/doi/10.1103/PhysRevLett.129.130402}
}

@Article{Goldman2016,
	author={Goldman, N.
	and Budich, J. C.
	and Zoller, P.},
	title={{Topological quantum matter with ultracold gases in optical lattices}},
	journal={Nature Physics},
	year={2016},
	month={Jul},
	day={01},
	volume={12},
	number={7},
	pages={639-645},
	issn={1745-2481},
	doi={10.1038/nphys3803},
	url={https://doi.org/10.1038/nphys3803}
}

@article{Osterloh2005,
	title = {{Cold Atoms in Non-Abelian Gauge Potentials: From the Hofstadter "Moth" to Lattice Gauge Theory}},
	author = {Osterloh, K. and Baig, M. and Santos, L. and Zoller, P. and Lewenstein, M.},
	journal = {Phys. Rev. Lett.},
	volume = {95},
	issue = {1},
	pages = {010403},
	numpages = {4},
	year = {2005},
	month = {Jun},
	publisher = {American Physical Society},
	doi = {10.1103/PhysRevLett.95.010403},
	url = {https://link.aps.org/doi/10.1103/PhysRevLett.95.010403}
}

@article{Goldman2009,
	title = {{Non-Abelian Optical Lattices: Anomalous Quantum Hall Effect and Dirac Fermions}},
	author = {Goldman, N. and Kubasiak, A. and Bermudez, A. and Gaspard, P. and Lewenstein, M. and Martin-Delgado, M. A.},
	journal = {Phys. Rev. Lett.},
	volume = {103},
	issue = {3},
	pages = {035301},
	numpages = {4},
	year = {2009},
	month = {Jul},
	publisher = {American Physical Society},
	doi = {10.1103/PhysRevLett.103.035301},
	url = {https://link.aps.org/doi/10.1103/PhysRevLett.103.035301}
}

\end{document}